\documentclass[fleqn,10pt]{wlscirep}
\usepackage[utf8]{inputenc}
\usepackage[T1]{fontenc}

\newcommand{\disNum}{D}
\newcommand{\disTel}{p}
\newcommand{\disLoc}{\lambda}
\newcommand{\pUp}{I}

\title{Network location and clustering of genetic mutations determine chronicity in a stylized model of genetic diseases}

\author[1,2]{Piotr Nyczka}

\author[2,*]{Johannes Falk}

\author[2]{Marc-Thorsten H\"utt}

\affil[1]{Faculty of Management, Wrocław University of Science and Technology}
\affil[2]{Department of Life Sciences and Chemistry, Jacobs University, D-28759 Bremen, Germany}

\affil[*]{j.falk@jacobs-university.de}

\begin{abstract}
In a highly simplified view, a disease can be seen as the phenotype emerging from the interplay of genetic predisposition and fluctuating environmental stimuli. We formalize this situation in a minimal model, where a network (representing cellular regulation) serves as an interface between an input layer (representing environment) and an output layer (representing functional phenotype). Genetic predisposition for a disease is represented as a loss of function of some network nodes. Reduced, but non-zero, output indicates disease. 
The simplicity of this genetic disease model and its deep relationship to percolation theory allows us to understand the interplay between disease, network topology and the location and clusters of affected network nodes. We find that our model generates two different characteristics of diseases, which can be interpreted as chronic and acute diseases. In its stylized form, our model provides a new view on the relationship between genetic mutations and the type and severity of a disease.
\end{abstract}

\begin{document}

\flushbottom
\maketitle
%
%
\thispagestyle{empty}

\section*{Introduction}
\label{sec:intro}

The debate on how to define disease is shaped by the necessity of balancing formal definitions motivated by physiology and correspondence with societal norms and expectations~\cite{merskey1986variable,margolis1986thoughts,cooper2002disease}. At the core of early definitions of diseases is a dysfunction of an organismal subsystem (often on the systemic level of organs), which affects the evolutionary goals of the organism as a whole~\cite{ereshefsky2009defining,pearce2011disease}.  The challenge of defining disease is also reflected in the multitude of disease ontologies~\cite{haendel2018census} and the limited ability to create mappings among them~\cite{harrow2017matching}. 

On a theoretical level, the multi-stage model of carcinogenesis~\cite{nordling1953new,armitage1954age} is an early example of formalizing diseases using a high-level abstraction in terms of a mathematical framework by writing down an explicit equation for the incidence as a function of age, based on the assumption of carcinogenesis as a multi-stage process~\cite{armitage1954age}. The model has been extended by Rozhok and DeGregori~\cite{rozhok2015toward} including environmental factors and in~\cite{rozhok2019generalized} incorporating additional levels of detail, leading to insight into disease mechanisms and in particular, providing an explanation of the nearly universal age-dependent incidence patterns observed across many cancers. This evolutionary model of cancer considers oncogenic mutations as well as tumour microenvironment and tissue architecture. The question of, how in the case of cancer the environment contributes to risk has been addressed in~\cite{hochberg2017framework}, where the necessity of adopting an ecological perspective on diseases has also been pointed out. 

A recent review~\cite{liu2020computational} summarizing the application of network biology to human diseases illustrates how disease mutations can be thought of in a network and emphasizes the importance of considering biological networks embedded in an environmental context. 

As an illustration of this avenue of research to one specific non-cancer disease, in~\cite{victor2016network} the authors have created a modular graph model to describe incidence curves for Crohn's disease, a disease currently in the focus of interest of Systems Medicine~\cite{knecht2016distinct,bauer2017interdisciplinary,hasler2017uncoupling,fiocchi2021ibd}. 

Most approaches in Systems Medicine fall into two categories, (1) employing mathematical or computational concepts to analyze medical data and (2) employing mathematical or computational concepts to model a specific disease or class of diseases. 
In contrast to these data-driven or single-disease approaches, we here strive for a model-driven understanding of some generic relationships between environment, genotype and disease phenotype. To this end, we distill the diverse concepts into a highly stylized model of an abstract genetic disease. 
A suitable framework is a complex system  ${\cal C}$ receiving at each moment in time $t$ an input vector  ${\cal I}(t)$ (representing environmental stimuli) and generating an output vector  ${\cal O}(t)$ indicating systemic function. 

Our model allows us to simulate the interaction of the disease (represented as a loss of function of some network nodes) and a fluctuating environment (represented by inputs to the network). The simplicity of the model enables us to investigate in detail how the observed features (disease severity, incidence curves, etc.) depend on the topology of the network and on the characteristic of the disease.

An important component of the model is that it operates using Boolean logic: Binary inputs are processed via Boolean ANDs (representing complete dependence) and ORs (representing the possibility of choice) and yield a binary output vector. Due to its minimal character and design, we are able to map the concept of directed percolation~\cite{broadbentPercolationProcessesCrystals1957} onto our disease model. Our model, therefore, allows harnessing the extensive knowledge of statistical physics about percolation phenomena~\cite{hinrichsenNonequilibriumCriticalPhenomena2000,hinrichsenPossibleExperimentalRealizations2000} for the analysis of diseases. 

Detailed analysis of our model suggests the following core properties: \textit{Chronic diseases} occur predominantly, when clusters of affected nodes are proximal to the output layer (representing network function or phenotype) and are enhanced by network connectivity (higher branching). \textit{Acute diseases}  tend to be independent of the position of affected nodes in the network. Higher branching transforms acute diseases into chronic diseases, but also in general reduces the likelihood of disease.

We further find that for a high number of OR nodes high connectivity between pathways mitigates the severity of a disease. In contrast, for a high number of AND nodes, low connectivity mitigates the severity. Additionally, we find that the impact of the position of the disease-affected nodes increases with the connectivity and decreases with the fraction of AND nodes.

\section*{\label{sec:methods}Methods}
Our disease model is motivated in parts by genome-scale metabolic models \cite{terzer2009genome,o2015using} and flux-balance analysis \cite{kauffman2003advances,orth2010flux}, where nutrient availability and the choice of the cellular objective function (e.g., maximization of growth or energy output) determine the steady-state pattern of metabolic fluxes. 
It also bears similarity to random Boolean networks as minimal models of gene-regulatory systems, where discrete time and the reduced state space allow for an analysis of attractors and their robustness \cite{kauffman1969metabolic,bornholdt2005less}. As such, our model is in the tradition of minimal models (or 'toy models', 'stylized models') in statistical physics \cite{radde2016physics,sneppen2017models}. 

Figure \ref{fig:block_scheme} summarizes the general scheme of our investigation (Fig. \ref{fig:block_scheme}a), the formal definition of node states (Fig. \ref{fig:block_scheme}b), the notion and dynamical effect of branching (Fig. \ref{fig:block_scheme}c) and the layer structure of the model (Fig. \ref{fig:block_scheme}d). In addition to obvious size parameters (number of input nodes, number of layers), our model depends on two parameters, the branching probability $b$ and the ratio $a$ that defines the fraction of nodes that act as AND or OR gates. 

\subsection*{Disordered lattice model}

\begin{figure*}[t!]
	\centering
	\includegraphics[width=0.9\linewidth]{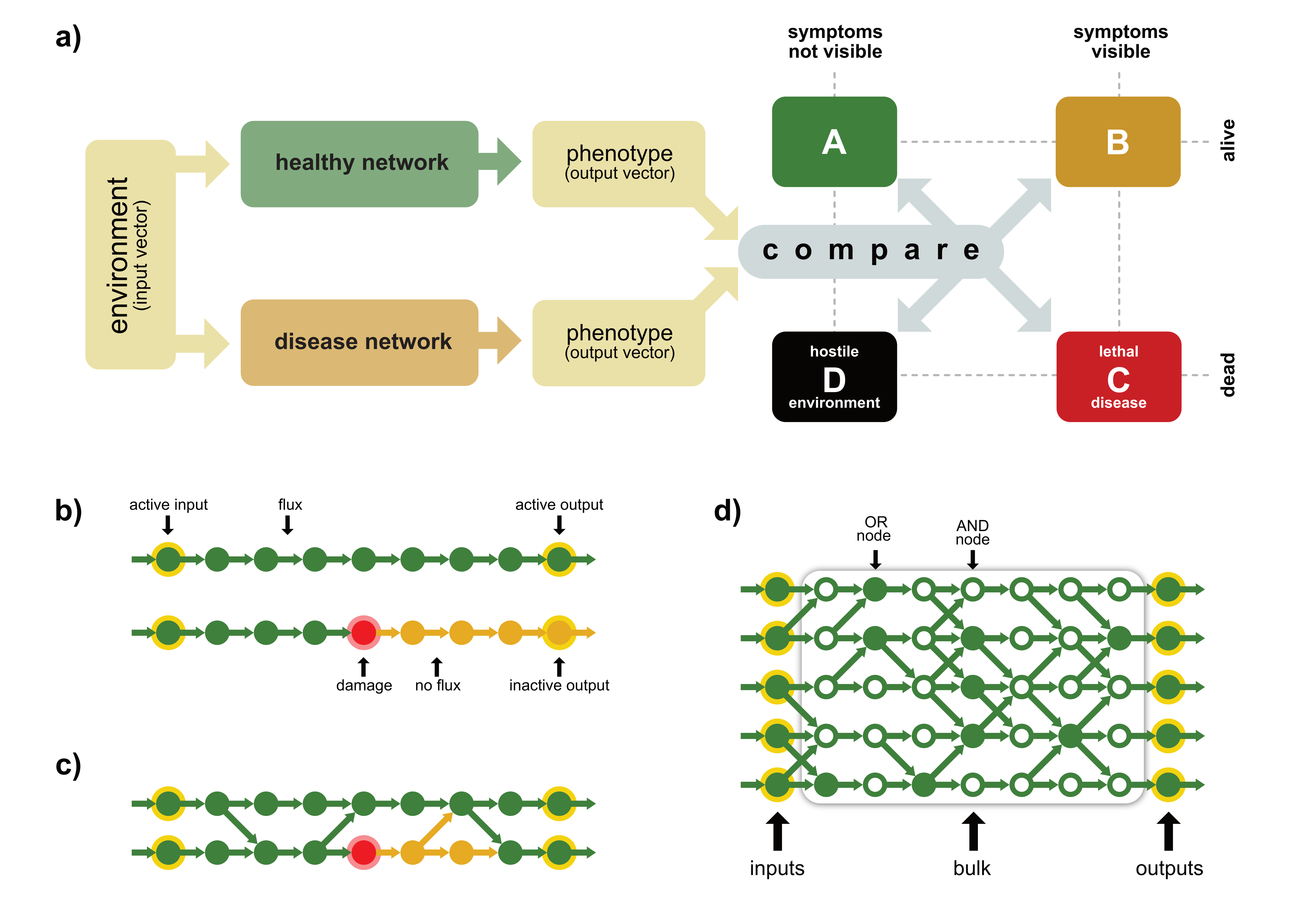}
	\caption[]{\label{fig:block_scheme} (a): Flow chart of our investigation: The outputs of a healthy and a defect (disease) network (receiving the same input vector) are compared. Depending on their difference they are assigned to a specific class: Class A, if both outputs are equal (no disease), class B, if the disease network has a lower but non-zero output (symptomatic disease), class C, if the disease network has zero output, while the healthy network has a non-zero output (lethal disease), class D, if both the healthy and the disease network have zero output.\\ 
	(b): Definition of node states and visual representation of the effect of a non-functional node. (c): Illustration of branching and visual representation of compensating a defect via interacting pathways. (d): Schematic representation of the full model and illustration of the layer structure.}
\end{figure*}

Motivated by biological network maps~\cite{barabasiNetworkMedicineNetworkbased2011a,gohHumanDiseaseNetwork2007} we write the generic biological network representing genotype and interfacing environment (input layer) and phenotype (output layer) as a set of $H$ parallel pathways (as depicted in Fig.~\ref{fig:block_scheme}d). Each idealized pathway is represented as a directed line graph with $L$ nodes, where each node represents a functional unit of the network that can either be active or inactive. We characterize each node by its pathway index $k \in \{1 \dots H\}$ and its position $j \in \{1 \dots L\}$ in the pathway (number of steps from input towards output). Following the definition of a line graph, each node $(k,j)$ is hence connected to the following node within the same pathway $(k,j+1)$ via a directed edge. To incorporate generic dependencies between different pathways (e.g. regulatory and discriminating mechanisms or regulatory overlap of metabolic pathways) with a probability given by the branching parameter $b$, a node $(k,j)$ is connected to the following node of a neighboring pathway $(k-1,j+1)$ or $(k+1,j+1)$, respectively. We do not employ any type of periodic boundary conditions. Hence, connections outside the obvious boundaries are omitted. Due to this branching, each node can have up to three inputs. For simplicity,  we assume that the processing performed at each node is represented by one of two possible Boolean functions (logical AND or logical OR) determining the local input-output relation of this node. The parameter $a$ determines the ratio between the number of nodes that act as ANDs (and consequently $1-a$ is the percentage of ORs).

The environment is represented by presences and absences of input components ('stimuli', 'nutrients') and hence by a binary vector. This input vector, together with the processing capabilities of each node, then creates a flux pattern of active nodes and links which finally results in an output vector.

This model of interdependent pathways is, of course, only a stylized approximation of a real-life biological network. In order to keep the interactions as simple as possible, the model is based on several idealizations: 
(1) Due to the enforced lattice structure, our model assumes that within the network only neighbouring pathways can be connected. This does not reflect the topology of real networks. However, for the sake of phenomenological insight into the local correlations between pathways, we decided to concentrate on this lattice structure. (2) As an acyclic graph our model does not allow for loops. This is at odds with the fact that regulatory elements heavily rely on direct or indirect feedback mechanisms. Feedback loops are often associated with rapid systemic responses to perturbations~\cite{alonNetworkMotifsTheory2007,doncicFeedforwardRegulationEnsures2013}. In this sense, our simplifying assumption is comparable to a steady-state approximation (e.g. employed in flux-balance analysis in metabolic investigations~\cite{orth2010flux,varmaMetabolicFluxBalancing1994}). Also, note that the nodes in our network represent functional units that might internally rely on different feedback structures. 

These idealizations do not allow for a one-to-one mapping onto real biological systems. Our disordered lattice functions as an \textit{effective network} summarizing the joint action of a multitude of biological networks -- from signalling pathways~\cite{katohWNTSignalingPathway2007,guptaBooleanNetworkAnalysis2007} and protein interactions~\cite{vazquezGlobalProteinFunction2003,vazquezModelingProteinInteraction2003} to metabolic networks~\cite{christensenMetabolicNetworkAnalysis1999,sungGlobalMetabolicInteraction2017}.

\begin{figure*}[t!]
	\centering
	\includegraphics[width=.9\linewidth]{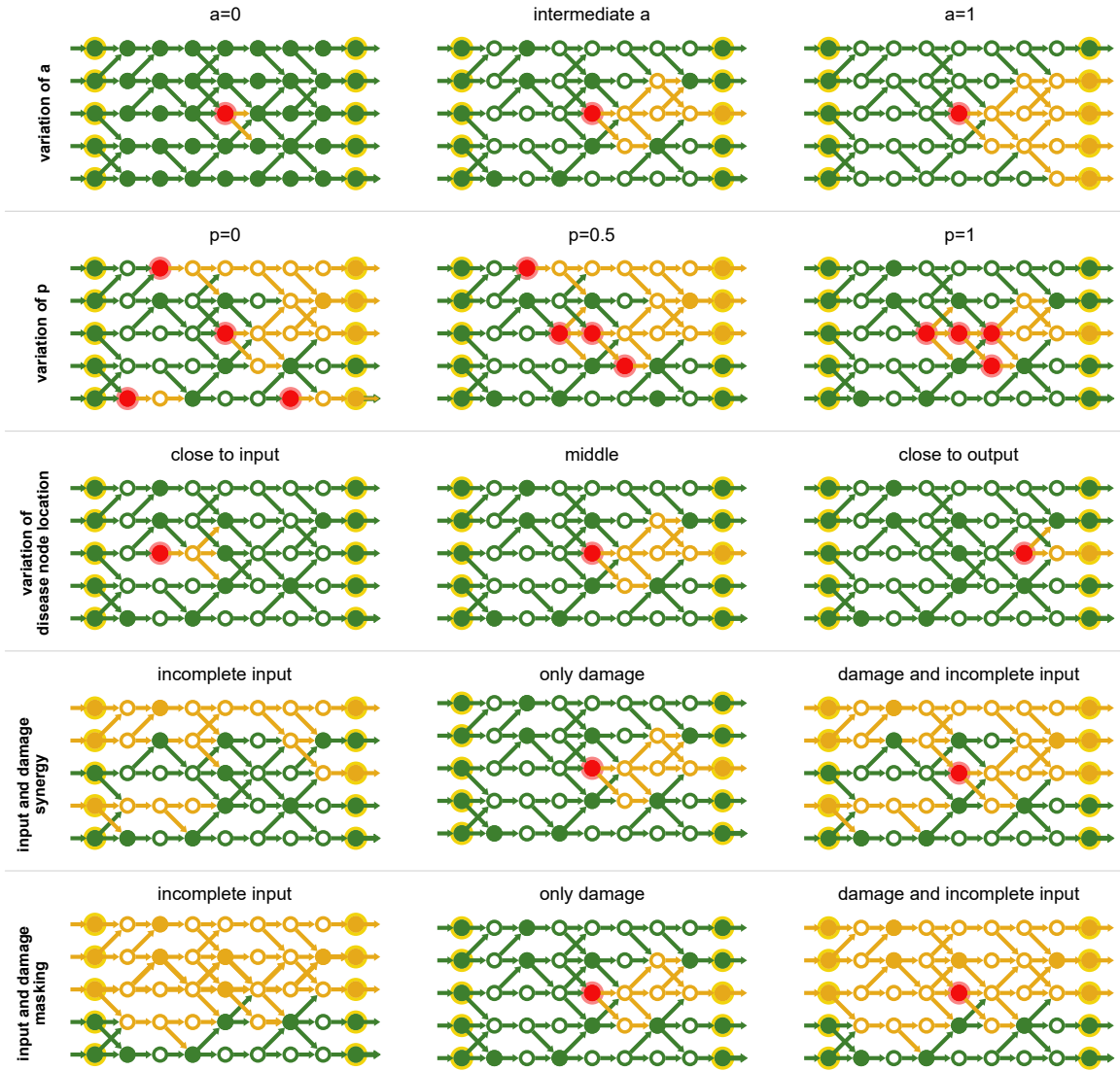}
	\caption[]{\label{fig:lattice_disease_example} Schematic summaries of various aspects of the model. Red indicates non-functional nodes (i.e. the genetic predisposition for a disease, `disease nodes'). Green nodes and links indicate activity (`flux'). Inactive components are indicated in yellow. All rows (except for the first one) use the same (intermediate) value of $a$. The first row shows how $a$ affects the impact of non-functional nodes. The second row illustrates the effect of different clustering $p$ for the same number of non-functional nodes $\disNum$. The third row shows how the position of the non-functional nodes can affect the phenotype. The fourth row is an example of synergistic effects between network and environment. The fifth row shows an example of a genetic disease being masked by environmental factors.}
\end{figure*}

For biological systems, where the knowledge about interactions of biological entities is more complete than for human cells, e.g., bacteria, it has been shown that the precise interplay of genetic regulation and metabolism installs a balance between robustness to environmental fluctuations and sensitivity to genetic changes~\cite{grimbsSystemwideNetworkReconstruction2019,klosikInterdependentNetworkGene2017,sonnenscheinAnalogRegulationMetabolic2011}. By adjusting the ratio between AND and OR gates we can continuously tune our model to such a robust or sensitive behaviour. For example, in the limit of only AND nodes ($a=1$), a single deactivated input will cause the deactivation of any connected downstream node. Likewise, in the limit of only OR nodes ($a=0$), a single activated input activates all connected downstream nodes. The first row of Fig.~\ref{fig:lattice_disease_example} illustrates this effect. 

\subsection*{Disease generation}

As with many other examples of minimal models or 'toy models' in biology (see \cite{falkMinimalModelBurstnoise2017a,kosmidisMinimalModelGene2021,radde2016physics}), the stylized nature of our model allows us to formally represent many features of a real-life system. In the following, we seek to study basic disease characteristics.

Within our model, a genetic disease -- a genetic predisposition for a disease phenotype --  manifests as a loss of function of one or several network nodes as shown in Fig.~\ref{fig:block_scheme}b. We characterize such a disease by a set of properties: the number $\disNum$ of disease-associated (defect) nodes and their distribution, determined by the clustering parameter $\disTel$ and the resulting average location $\disLoc$. 
To define which nodes are affected, the first node is chosen randomly. Then, the selection of the $\disNum -1$ remaining nodes relies on the Eden growth model~\cite{edenTwodimensionalGrowthProcess1961,lambiotteRankingClusteringNodes2012,nyczkaInferringPatternGenerators2021} with teleportation and proceeds as follows:
\begin{itemize}
    \item With probability $\disTel$ a node connected to the current cluster of disease-affected nodes gets deactivated (growth of the current cluster)
    \item Otherwise (with probability $1-\disTel$) a randomly selected node gets deactivated and serves as a nucleus for a new cluster. 
\end{itemize}

The parameter $\disTel$ can hence be used to tune the model between a state, where the disease-associated nodes are either distributed randomly ($\disTel = 0$) or concentrated in one connected cluster ($\disTel = 1$) as illustrated in the second row of Fig.~\ref{fig:lattice_disease_example}.

\section*{\label{sec:results}Results}

\subsection*{Incidence curves}

We are now in a position to analyze how fluctuations of the environment affect an unperturbed ('healthy') network in contrast to a network with non-functional nodes representing a disease genotype.

Due to the focus on only AND or OR gates, a node can only become active, if at least one input was active. Now, since a disease appears as a deactivated node, for the same input a defect network has always the same or fewer active outputs than a healthy network: The active outputs of the disease affected network are always a proper subset of the active outputs of the healthy network.

 For our quantitative analysis, we first generate a random environmental condition (characterized by the probability $\pUp$ of active inputs). For every time step we then proceed as follows (compare Fig.~\ref{fig:block_scheme}a):
\begin{itemize}
    \item For the given environment we analyze the output vector of the healthy network. If the vector is zero, the environment is already lethal for a healthy and hence also for the defect network (case D in Fig.~\ref{fig:block_scheme}). 
    \item If the output of the healthy system is non-zero we compare its output to the output of a defect network (receiving the same input vector). There are three possible outcomes: (1) Both vectors are equal. This can be interpreted as no disease symptoms: Both networks display a healthy phenotype (case A in Fig.~\ref{fig:block_scheme}). (2) The vector of the defect network is non-zero but has fewer non-zero components than the healthy network. This case represents a disease phenotype (case B in Fig.~\ref{fig:block_scheme}). (3). The output vector of the defect network is zero. This indicates lethality due to the disease (case C in Fig.~\ref{fig:block_scheme}).
    \item To simulate fluctuations in the environment at each time step each element of the input vector is preselected for change with a fixed probability of 20\%. Then, each element within this preselected group is set to $1$ with probability $\pUp$ and to $0$ otherwise. 
    
\end{itemize}

\begin{figure*}[t]
	\centering
	\includegraphics[width=1.0\linewidth]{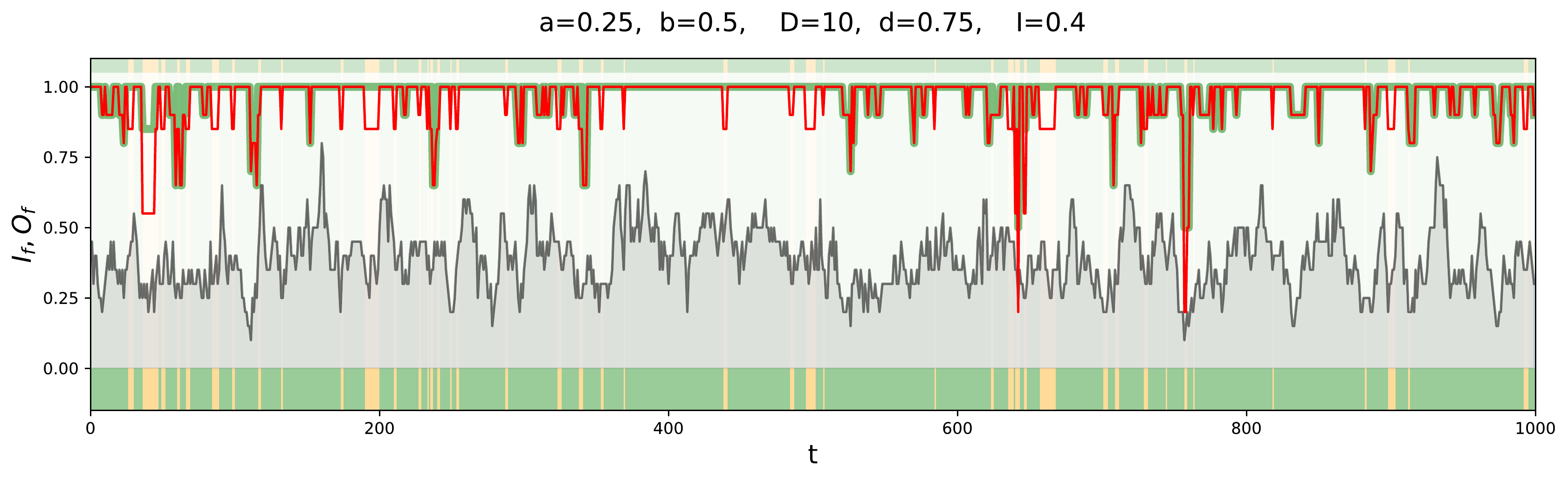}
	
	\hbox{\hspace{0.2cm}
	\includegraphics[width=0.97\linewidth]{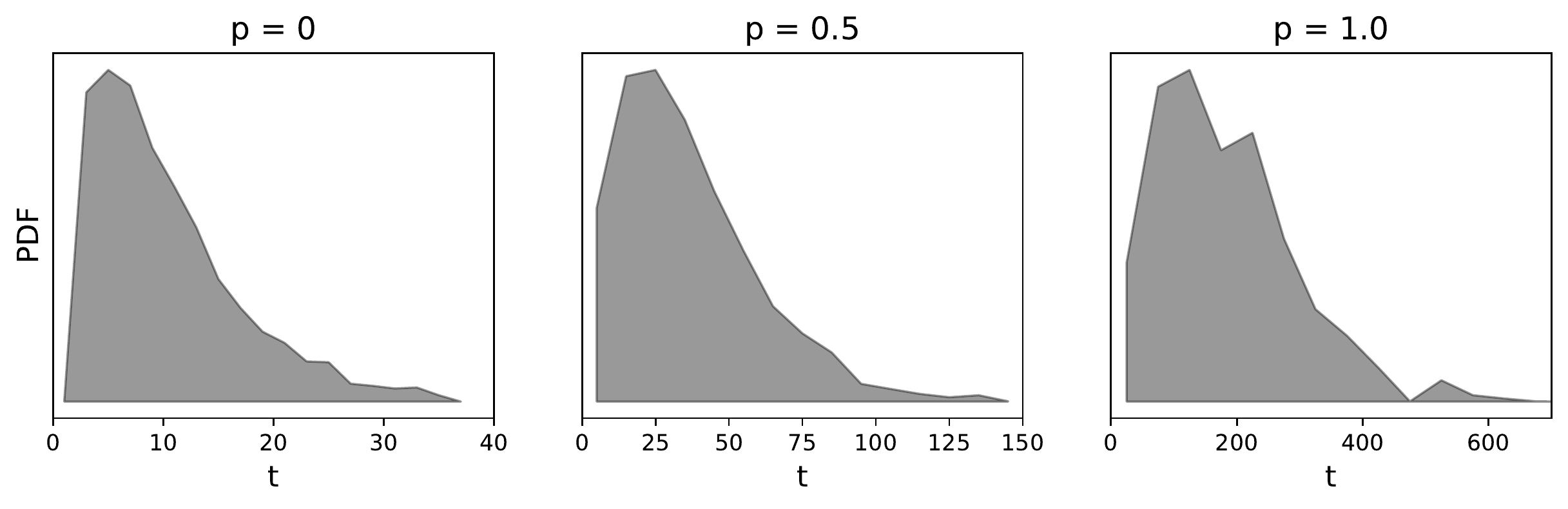}}

	\caption[]{\label{fig:incidence_curve} Top panel: time series of the number of active inputs (grey background) and the output strength for the healthy (green) and the disease (red) network. A difference between the red and the green curve (red below green) indicates visible symptoms. The type of the observed disease is indicated by the colour bar (colour code as defined in Fig. \ref{fig:block_scheme}a). Further time series for different sets of parameters can be found in Figs.~\ref{fig:trajectories_supplement_1} and~\ref{fig:trajectories_supplement_2}. Bottom panel: histograms of disease incidences for different values of $p$.}
\end{figure*}

As a result, we obtain time series as shown in Figures~\ref{fig:incidence_curve} (top). Here, the green and red curves represent the output strength of the healthy and the disease network, respectively. Every difference between the green and the red curves thus indicates a disease phenotype for the respective environment. The corresponding case of the four possible (no detectable disease, disease phenotype, death due to disease, death due to environmental conditions) is indicated by the respective colour in the bars in the lower segment. For all time series without death, it is possible to analyze the distribution of time spans with and without symptoms, which results in incidence curves as e.g. shown in Fig.~\ref{fig:incidence_curve} (bottom). 
Further time series are shown in Figs.~\ref{fig:trajectories_supplement_1} and~\ref{fig:trajectories_supplement_2}.

One should note that the environment characteristic can in principle be used to scale the incidence curve in time for comparison with a specific disease, e.g. by decreasing the frequency of the fluctuations. However, as we are interested in the universal features of the model we do not pursue this line of investigation here. 
We have shown that our model, despite its simplicity, is capable of producing realistic incidence curves (Fig.~\ref{fig:incidence_curve} (bottom)). In the following, we will now investigate how the different parameters influence the behaviour of the system.

\begin{figure*}[t!]
	\centering
	\includegraphics[width=0.9\linewidth]{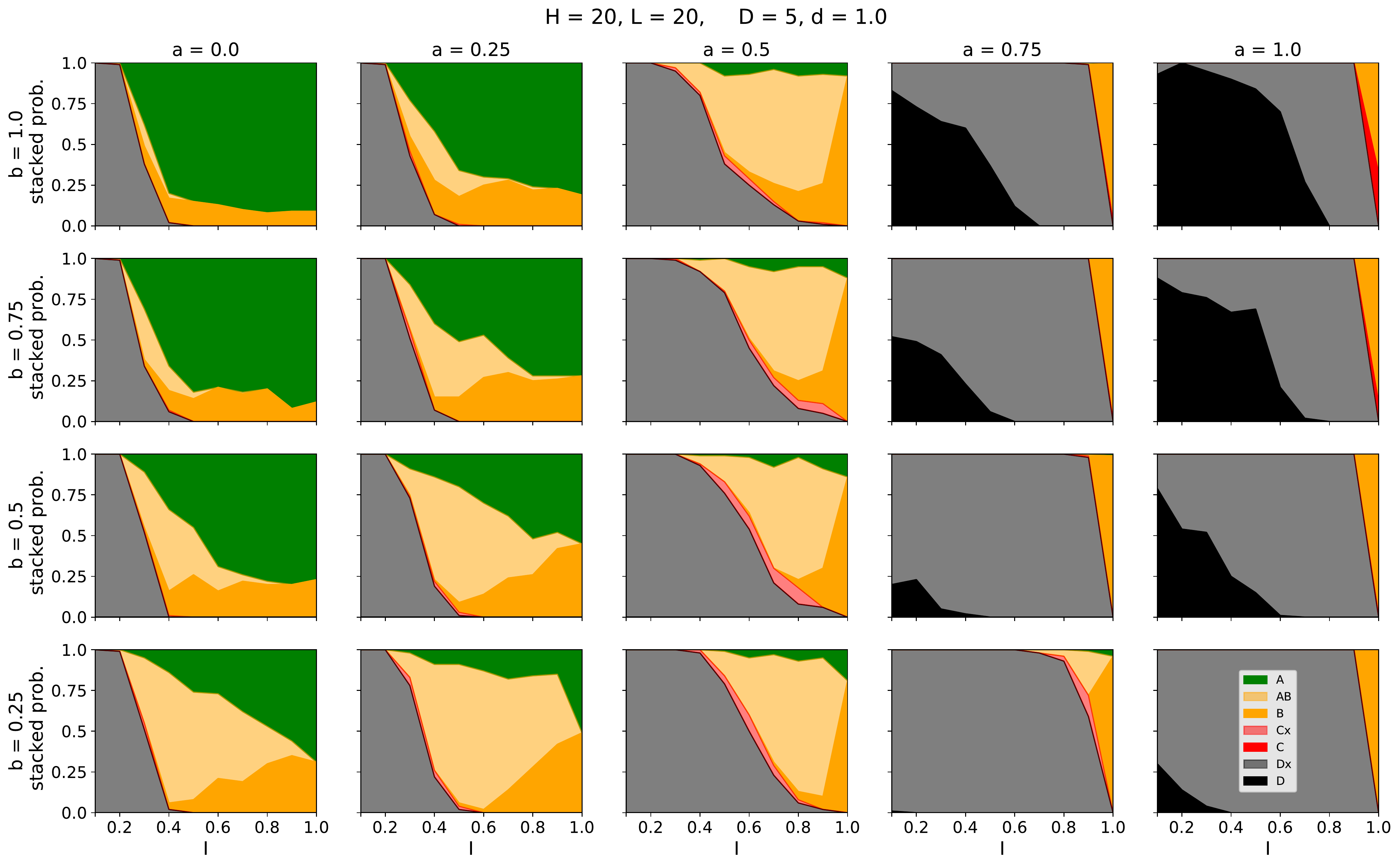}
	\includegraphics[width=0.9\linewidth]{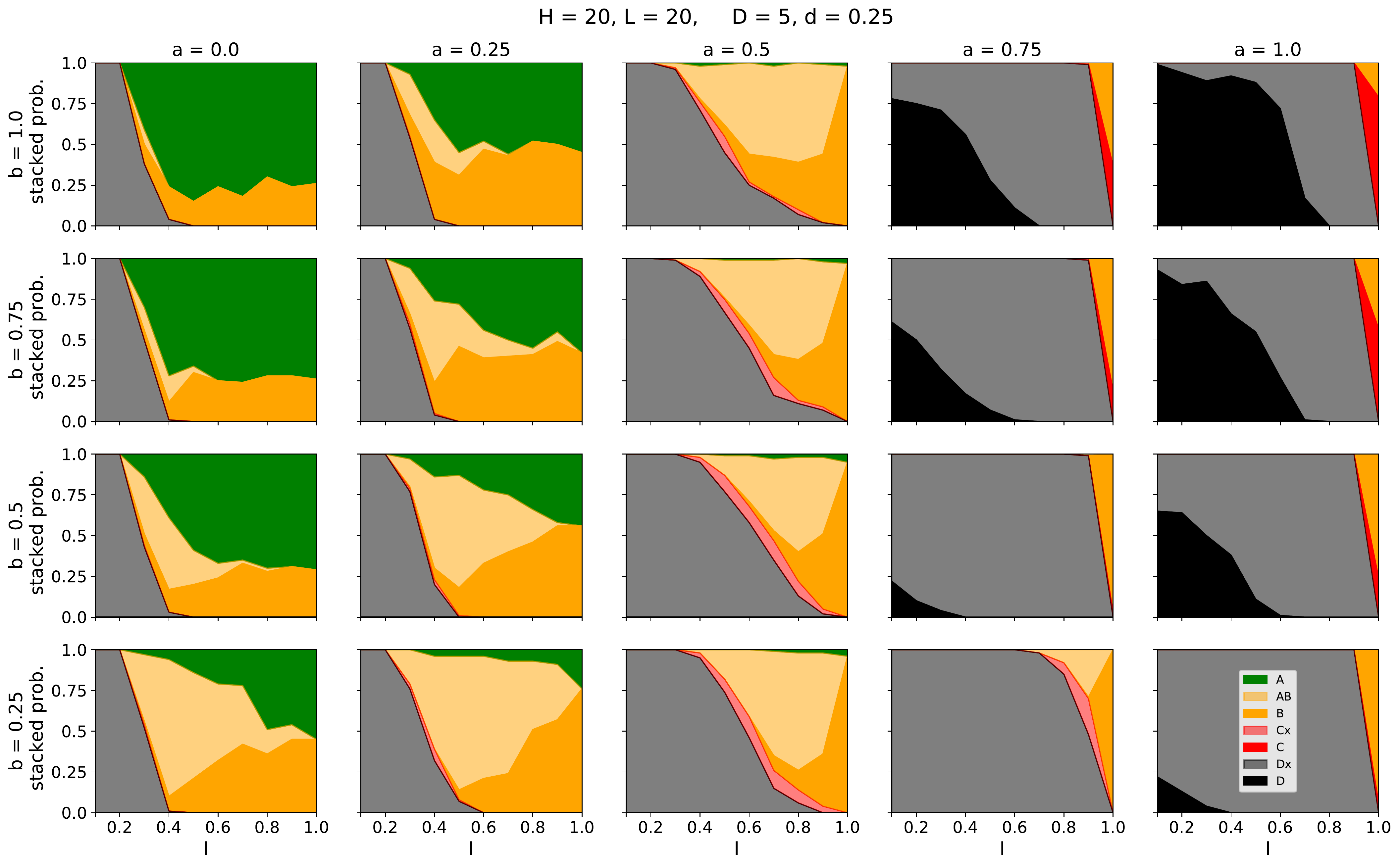}
	\caption[]{\label{fig:classification_1} Dependence of the observed cases on the fraction of active inputs $\pUp$. For each parameter combination, the system was simulated 100 times for 1000 time steps. The colour code (as defined in Fig.~\ref{fig:block_scheme}a) indicates, which cases occurred in the time line. Here, Cx and Dx denote that the time line also contained the cases (A, B) or (A, B, C), respectively. Curves for other sets of parameters are provided in Figs.~\ref{fig:dependence_input_supplement_1} and~\ref{fig:dependence_input_supplement_2}.
	}
\end{figure*}

Whether and how often the time series show a specific case depends on the choice of parameters. In Fig.~\ref{fig:classification_1} we vary the fraction of active inputs $\pUp$ and analyze for 1000 steps how often a specific case was reached. The same plots for other parameter combinations are presented in Figs.~\ref{fig:dependence_input_supplement_1} and~\ref{fig:dependence_input_supplement_2}. For a high fraction of AND gates ($a > 0.5$) the healthy network is already very sensitive and often shows zero output if only a few input elements are deactivated (frequent occurrence of case D). These sensitive systems become more robust, if the connectivity between the pathways is decreased. This dependence stems from the fact that low connectivity is likely to isolate the consequences of a genetic defect, by restricting it to very few pathways, or even a single one. In the special case of $b=0$, the system is just a collection of independent single pathways. In such a case the parameter $a$ does not have an effect and the output is always the same as the input. In contrast, for a large number of OR gates ($a < 0.5$), there is a high chance that the healthy, as well as the defect system, have non-zero output. Within this regime, for high branching $b > 0.5$ the outputs of both systems are likely to be equal because a deactivated pathway gets healed by neighbouring pathways as depicted in Fig.~\ref{fig:block_scheme}c. For low branching $b < 0.5$ the defect network often shows symptoms. Depending on the proportion of AND and OR gates the disease network hence has an advantage if either the branching is high or low. The figures also allow for another observation: For a high number of active inputs there are -- depending on the disease -- mainly two possibilities: Either the system stays in case A (the healthy and the disease network show the same output), or it stays in case B (the disease network shows lower output). If the fraction of active inputs is decreased, it is also possible (as e.g. observed in Fig.~\ref{fig:incidence_curve} (top)) that the time series switches between cases A and B. We can identify these outcomes with two different disease conditions: If the system stays in case B this corresponds to a \textit{chronic disease} where a lower output persists. Contrarily, if we observe a switching between A and B, this corresponds to diseases \textit{observed} as \textit{acute}.

\begin{figure*}[t!]
\centering
\includegraphics[width=0.9\linewidth]{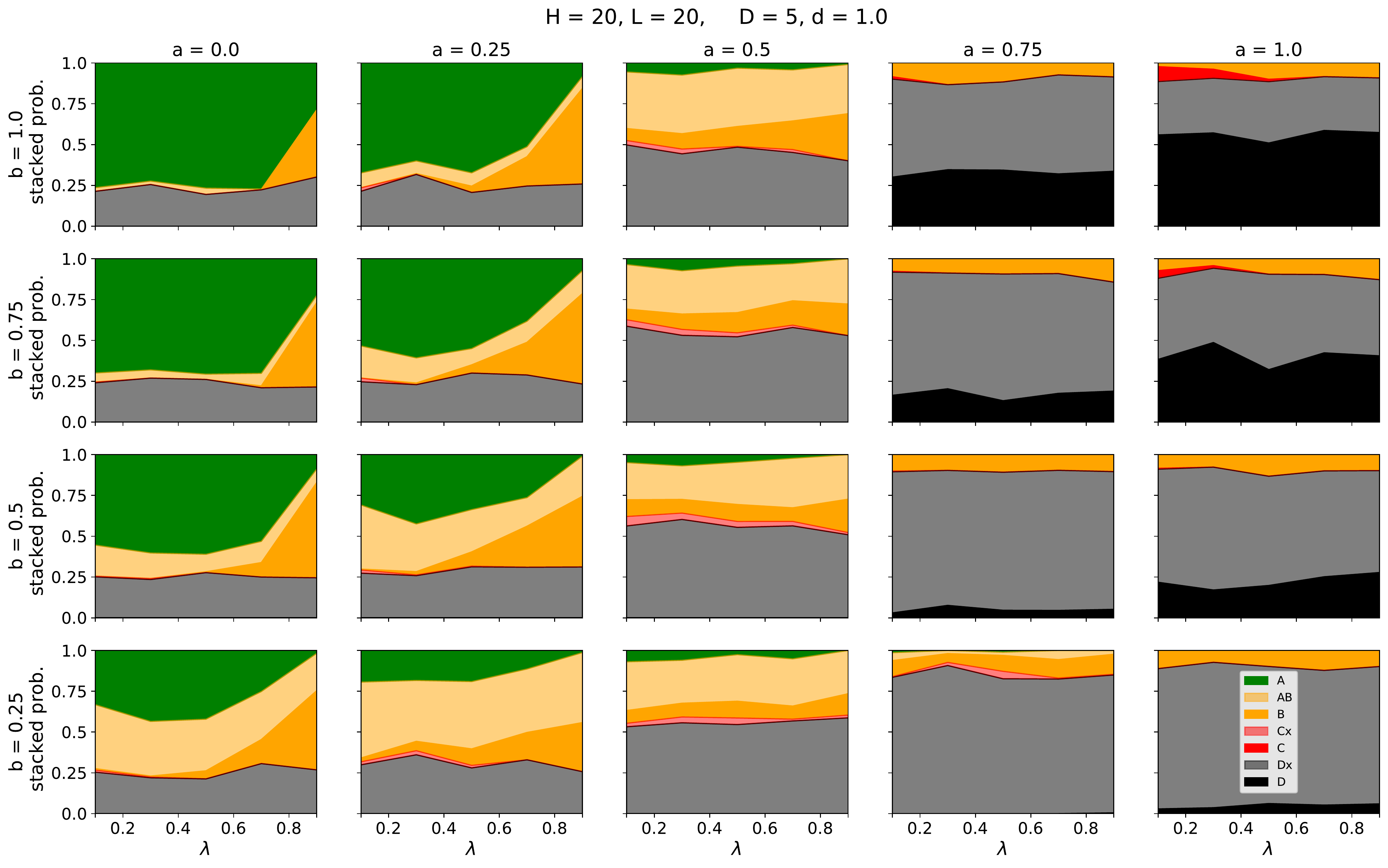}
\includegraphics[width=0.9\linewidth]{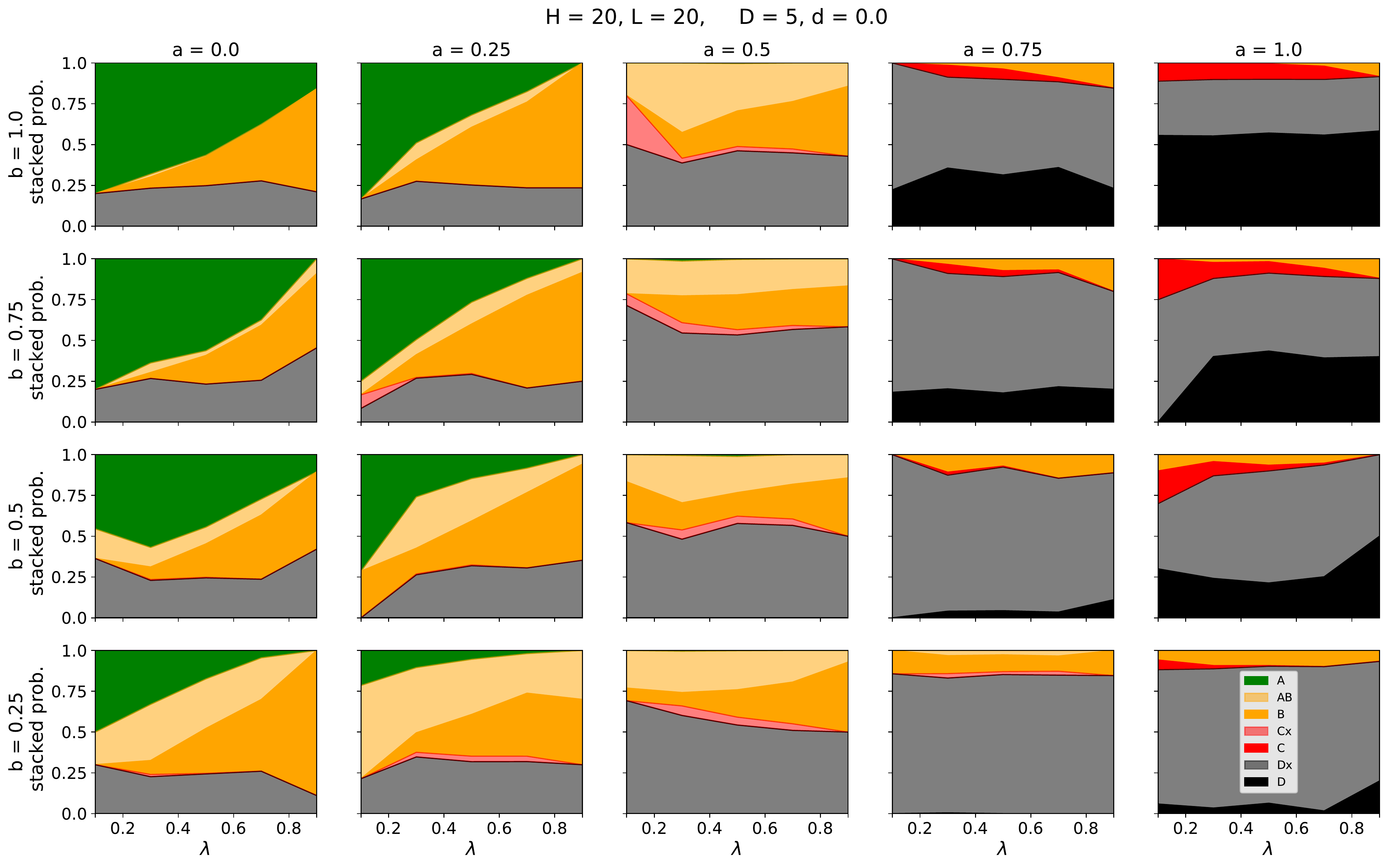}
	\caption[]{\label{fig:location_1} Dependence of the observed cases on the average position $\disLoc$ of inactive nodes. The colour code (as defined in Fig.~\ref{fig:block_scheme} a) indicates, which cases occurred in the time line. For the upper figure, the clustering parameter was set to $\disTel = 1$ which means that all defective nodes form one cluster. Contrarily, in the lower figure, the parameter was set to $\disTel = 0$, leading to a random distribution of the defective nodes. Results for other sets of parameters are provided in Figs.~\ref{fig:dependence_lambda_supplement_1} and~\ref{fig:dependence_lambda_supplement_2}.}
\end{figure*}

Besides the general dependence on the choice of parameters, it is particularly interesting to analyze, how the position of the disease nodes (within the network) affects the visibility of the disease.  Fig~\ref{fig:location_1} compares the possible outcomes of the network depending on the average location  $\disLoc$ of the disease-associated nodes as well on the fraction of ANDs $a$ and the branching parameter $b$ (Further results for different sets of parameters are presented in Figs.~\ref{fig:dependence_lambda_supplement_1} and~\ref{fig:dependence_lambda_supplement_2}).

For small $a$ we observe a strong dependence on the average location of the disease: If the average location is close to the input, the symptoms are often not visible and the output of the healthy and disease-affected network are equal (green, case A). Contrarily, if the average location is close to the output, the defect network has often less activity than the healthy network (the disease is visible; yellow, case B). 
For large $a$ the model shows different behaviour. Here, in most cases, both the healthy and the defect network have zero output (both are dead; black, case D). Additionally, the behaviour is mostly independent of the position of the non-functional nodes. We can explain this behaviour with some simple arguments: Let us assume a single non-functional node at location $k$ in the $j$th pathways. Without any crosstalk between the pathways, this defect affects all subsequent sites on position $k+1$, $k+2$, ... $L$. Now, if we allow for branching, two mechanisms need to be taken into account: (1) A signal from a neighbouring pathway ($j-1$ or $j+1$) can arrive and restart one of those affected nodes, which require a logical OR. (2) Since the transmission and distribution of 1s coming from the now deactivated pathway disappear, the single deactivated node can deactivate neighbouring pathways in case of a logical AND. Depending on the fraction of ANDs (determined by the parameter $a$) a disease affected node can hence create longer ``shadows'' of deactivated pathways or it can be circumvented. The branching determines the speed of these two mechanisms.

\subsection*{State-space dynamics}

The stylized nature of our model also allows for a more stringent and more comprehensive analysis, which is less based on numerical simulations, but on formalisms of discrete systems. This direction is pursued in the present section.

We consider the middle (bulk) segment of the network as an operator transforming the input vector (environment) into the output vector (phenotype). As the bulk is a set of consecutive layers, it is possible to trace the evolution of the input vector, step by step, all the way to the output. Using a state-space representation then allows us to analyze the evolution on the scale of the whole state space.
With $N$ input nodes (or parallel pathways) we formally have $2^N$ distinct input states. The corresponding $N$-digit binary numbers are then processed layer by layer. As this processing is deterministic, a single state at layer $k$ cannot give rise to multiple states at layer $k+1$. However, multiple states at layer $k$ can lead to the same state at layer $k+1$. Hence, as already described before, the diversity of states can only decrease across layers. This 'funnelling' of states along the network is instrumental for the functionality of our model.

\begin{figure}[t!]
	\begin{center}	
		\includegraphics[width=.8\textwidth]{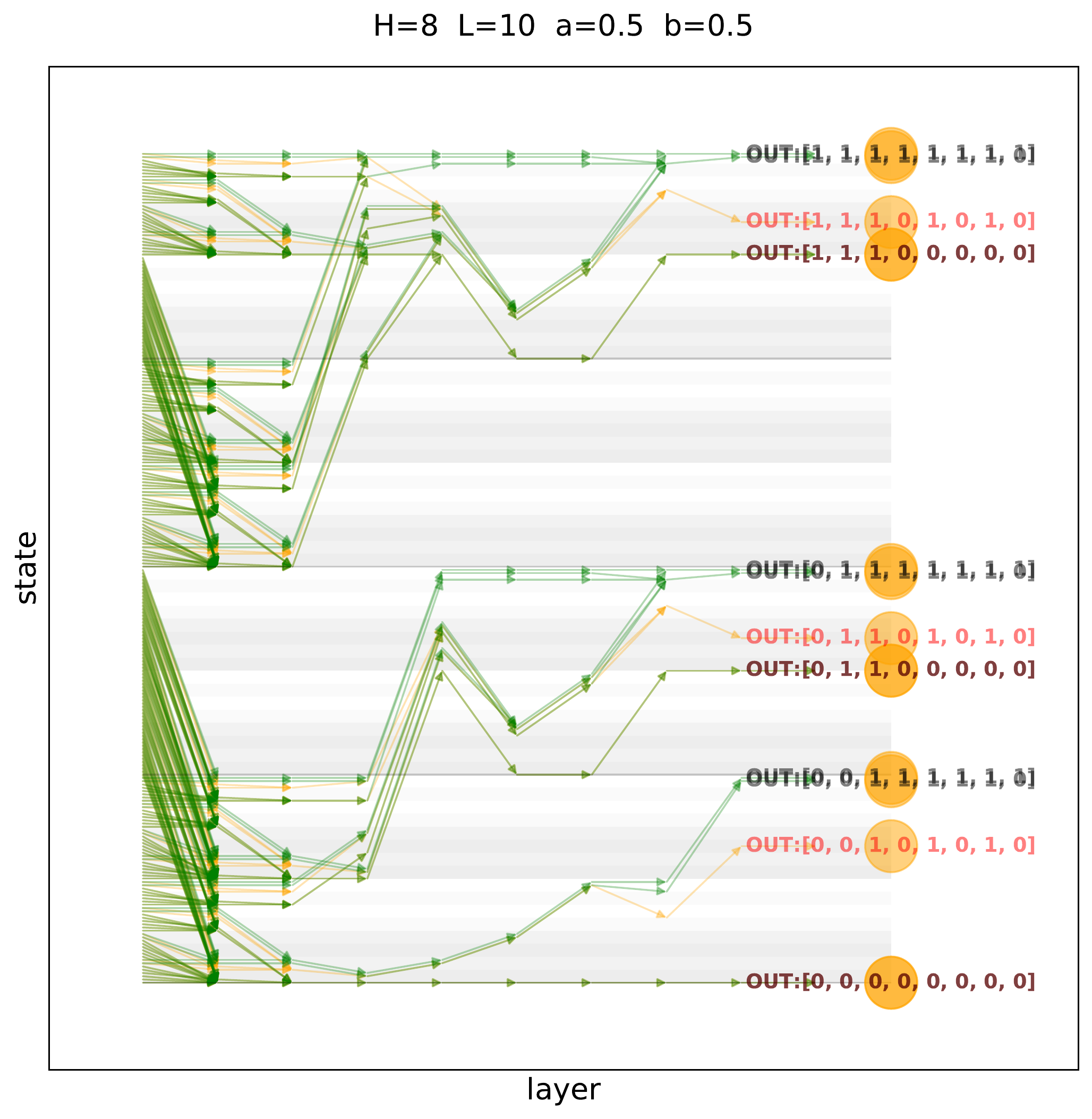}	
		\caption[]{Detailed view of the state space evolution. Input states (left; sorted according to the binary number they represent) are converted into output states (phenotypes; right) by the network. The number of phenotypes is usually much smaller than the number of possible input states. Green indicates a trajectory in the healthy network, while orange indicates a trajectory in the disease network.}
		\label{fig:evo_states_1}
	\end{center}
\end{figure}

The funnelling of binary states is illustrated in Fig.~\ref{fig:evo_states_1}, where all possible initial states are arranged along the y-axis and layers are shown along the x-axis, thus allowing us to follow all possible input states through the network. As soon as two lines meet they merge, which decreases the number of possible states after this time point by one. Formally, in the limit of infinite time (infinite length of the network) this always leads to a system where all outputs are either on or off.

\subsection*{\label{sec:percolation}Relation to percolation theory}

Biological regulatory networks can be classified as complex dynamical systems, the analysis of which has a long tradition in statistical physics~\cite{ladymanWhatComplexSystem2013}. The design of our model allows a specific class of models from physics -- \textit{directed percolation}(DP)~\cite{hinrichsenNonequilibriumCriticalPhenomena2000,broadbentPercolationProcessesCrystals1957} --  to be transferred to our system. More specifically, our model is similar to a subgroup of DP, namely \textit{Compact Directed Percolation}\cite{essamDirectedCompactPercolation1989a,domanyEquivalenceCellularAutomata1984,duarteSeriesMonteCarlo1992}. Percolation models are simple models from statistical physics to analyze signal propagation through heterogeneous systems e.g. cells or neurons~\cite{larkinSignalPercolationBacterial2018,zhouPercolationModelSensory2015}. In most of these models, a single parameter $p$ determines the probability that a signal is locally transmitted. If $p$ is too small, no signal can reach the output. In this (inactive) phase, the probability that the signal reaches a specific layer decays exponentially with the number of layers. Contrarily, if $p$ is large, there is almost always a connected path through the system and hence a finite probability that the signal reaches the output. The transition point between these two phases $p = p_c$ is, mathematically speaking, a critical point. At the critical point, the probability that the signal can traverse the medium decays algebraically with the number of layers. We can use these results from statistical physics to understand and interpret the results observed in our model. An example is Fig.~\ref{fig:location_1} where, from left to right, the fraction of ANDs was increased. At $a \approx 0.5$, we observe a sudden change in the general behaviour of the system: For $a \leq 0.5$ most systems show either case A or case B. However, for $a > 0.5$ most systems belong to case D where both systems show zero output. This transition can be identified as a phase transition within the genetic disease model.

Another central topic in the analysis of percolation models is their dependence on small perturbations. The analysis of how a single perturbation (often also called damage) evolves over time is known as damage spreading~\cite{herrmannDamageSpreading1990}. If one introduces damage and compares the difference to the unmodified network there are two possible results: The damage can spread, which ultimately leads to a system that evolves very differently from the original network, or the damage might disappear. We envision that the analysis of damage spreading transitions can be an interesting direction for further research within our disease model.  

\section*{\label{sec:discussion}Discussion and Conclusion}

We presented a minimal model to study the interplay between network topology and disease nodes. Our model can be used to analyze, how incidence curves and disease visibility depend on parameters like the clustering of disease genes or the cross-talk between pathways. The situation considered here is generic (i.e., not tailored to any specific set of biological processes). The model is motivated by the general properties of metabolic networks, where the most obvious type of environmental fluctuation is a change in nutrient availability. The output vector can be thought of as some type of cellular objective function (for example growth or energy production) as typically employed in genome-scale metabolic models, for example for flux-balance analysis~\cite{orth2010flux}. A design concept of our model is that genetic predisposition manifests as a loss of function,  which is a suitable model, if the signal processing does not include a logical NOT. Following this choice, we only employ logical ORs and ANDs. However, if one relaxes these constraints, there are obviously other possible choices for logical gates e.g. the functionally complete sets \{AND, NOT\} or \{NOR\}.

The conceptual foundation of our model is the basic fact that human diseases are rarely the consequence of a single defective gene, but the result of complex interactions within the cellular-molecular network~\cite{barabasiNetworkMedicineNetworkbased2011}. The disease phenotype is hence a result of different and mutually dependent interactions. 

Although there are successful attempts to identify disease genes, due to the complexity of medical situations, the typically low signal strength, as well as the small sizes and high intrinsic diversity of disease cohorts, these approaches so far often provide limited functional insight. We believe that minimal, generative models of typical data types, as well as stylized representations of typical medical scenarios, are necessary to organize the analysis of the intricate relationship between genetic risk factors, environmental stimuli and observed disease phenotype. Our model allows building such an understanding from a general point of view: By the variation of a few parameters, it is possible to compare the interplay of different network topologies and disease characteristics. We also show that the specific pattern of a genetic predisposition of a biological network can have a direct and systematic impact on the disease phenotype. Such relationships between genetic defect patterns and phenotype patterns are a direct consequence of the architecture of the underlying network.

Our model stratifies diseases according to four main model properties: (1) high or low clustering of affected nodes (representing genetic predisposition), (2) strong or weak network connectivity (branching), (3) high or low numbers of ORs (regulatory alternatives) vs. ANDs (regulatory interactions), and (4) the clustering of affected nodes either proximal to input layer (representing environment) or proximal to the output layer (representing network function or phenotype) and thus the average position of affected nodes. 

Based on the detailed analysis of our model, we arrive at the following picture: High average position, high clustering and high branching facilitate chronic diseases. The average position of affected nodes does not strongly affect the probability of acute disease, in contrast to the clustering of these nodes, which disfavors acute diseases. 

Employing mathematical modelling to leverage biological networks -- and signalling networks in particular -- for the purpose of understanding human diseases is a cornerstone of the emerging field of precision medicine~\cite{yadavPrecisionMedicineNetworks2020,hastings2020applications}. Due to the simple structure of the network, our model allows for an in-depth and node-by-node analysis of the observed results. The model can hence be used to assess the robustness and vulnerability -- common topics in Systems Biology~\cite{kitanoComputationalSystemsBiology2002a,kitanoSystemsBiologyBrief2002a} -- of phenotypic states from a functional point of view.

\section*{Appendix: Evaluation of the parameter $a$}
\label{sec:appendix_A}
In order to provide some intuition on the range of values for the percentage $a$ of logical ANDs -- which is one of the key parameters of our model -- we resort to genome-scale metabolic models. Here we employ two strategies to estimate the value of $a$ from such metabolic reconstructions.

The first method counts the Boolean operations in the gene-to-reaction mappings within genome-scale metabolic models and computes the ratio
\begin{equation}
    a = \textrm{AND}/(\textrm{AND}+\textrm{OR}) .
\end{equation} 
This yields -- depending on the model -- $a$ between $0.105$ and $0.15$ (see Table \ref{tab:estimation_a}). The analysis was performed with the COBRA package for Python (Cobra version 0.19.0; Python version 3.6).

\begin{table}[h]
\caption{\label{tab:estimation_a} Estimation results for the parameter $a$ for three genome-scale metabolic models of human cells, Recon 1 \cite{duarte2007global}, Recon 2 \cite{thiele2013community} and Recon 3D \cite{brunk2018recon3d}.}
\begin{center}
\begin{tabular}{ |p{5.5cm}|p{2cm}|p{2cm}|p{2cm}|  }
	\hline
	\multicolumn{4}{|c|}{Estimation of $a$} \\
	\hline
	Method& Recon 1& Recon 2.3 &Recon 3D \\
	\hline
	1 (gene-reaction associations)  & 0.15 & 0.129 & 0.105 \\

	2 (complex reactions) & 0.24 & 0.08 & 0.21 \\
	\hline
\end{tabular}
\end{center}
\end{table}

The second method evaluates each (reaction) node together with its next-to-nearest neighbours. In this sense, the method is closer to the nature of nodes in our model, which do not represent individual reactions, but rather more complex regulatory entities, summarizing metabolic flow and genetic control. Such subgraph objects may behave like AND or OR gate depending on the mutual proportions of the alternative sources and the number of reactants.

This method utilizes information about reaction network including their directionality. Each reaction has a form of $R_1 + R_2 + ... \rightarrow P_1 + P_2 + ...$, where $R_i$ are reactants and $P_j$ products. Our quantification focuses on reactants alone. Each reactant may have one or more possible sources, where the number of these sources for each reactant is $k_i$. Hence, the reaction itself corresponds to a Boolean AND, but alternative sourcing is an analogue of a Boolean OR. Such a subgraph has the structure of multiple logical ORs fed to an single logical AND: $(R_{1,1} | R_{1,2} | ...) + (R_{2,1} | R_{2,2} | ...) + ... \rightarrow P_1 + P_2 + ...$. In order to estimate the parameter $a$, we need to convert this whole entity to just a single AND or OR.

All fluxes are regarded to be discrete, hence inputs can be characterized by a probability $c_{in}$ of having reactant $i$ from the source $j$, $R_{i,j}$. With this quantity,  the probability of reaction taking place, $c_{out}$, can be computed. For a logical AND $c_{out} < c_{in}$ whereas for a logical OR $c_{out} > c_{in}$, there are also special cases of $c_{in} = 0$ and $c_{in} = 1$. The same classification can be done for more complicated structures, like the one discussed above. For simple gates, there is always (for each $c_{in}$) either $c_{out} < c_{in}$ or $c_{out} < c_{in}$ depending on the gate. However, for more complicated entities this inequality may change sign upon change of $c_{in}$. Way to overcome this difficulty is to compare whole range of $c_{in} \in [0,1]$ by taking integral and then comparing: 
\begin{eqnarray}
	c_{out} &=& \prod_i (1-(1-c_{in})^{k_i}) \\
	\Delta_c &=& c_{out} - c_{in} \\
	\alpha &=& \int_0^1 \Delta_c dc_{in} = \int_0^1 \left (\prod_i (1-(1-c_{in})^{k_i}) - c_{in} \right )  dc_{in} , 
\end{eqnarray}
where $i$ is the metabolite index and $k_i$ is the number of alternative sources for this metabolite. Reactions with empty sets of sources were removed from the analysis.

Then if $\alpha > 0$ node is regarded as OR, in the case of $\alpha < 0$ node is AND. The value of $a$ obtained with this method is between 0.08 and 0.24 for human metabolic models (see Table \ref{tab:estimation_a}).
These values are clearly located in the subcritical regime. This is in line with our model results, which show that only the subcritical regime is viable and resilient to defect and perturbations.

We also checked metabolic models for other living organisms and we found substantial agreement between the values of $a$. This is a suitable starting point for further investigation. Also, the precise determination of this parameter deserves a more detailed study.

\section*{Author contributions}
MH and PN conceived this study. PN and JF performed numerical investigations. MH, PN and JF wrote the manuscript. 

\clearpage
\pagebreak





\bibliography{minimal_disease_model_medical_implications_v1.bib}






\clearpage
\newpage
\section*{Supplementary information}

\setcounter{figure}{0}
\renewcommand{\thefigure}{S\arabic{figure}}

\setcounter{table}{0}
\renewcommand{\thetable}{S\arabic{table}}

\begin{figure*}[ht]
	\centering
	\includegraphics[width=1.0\linewidth]{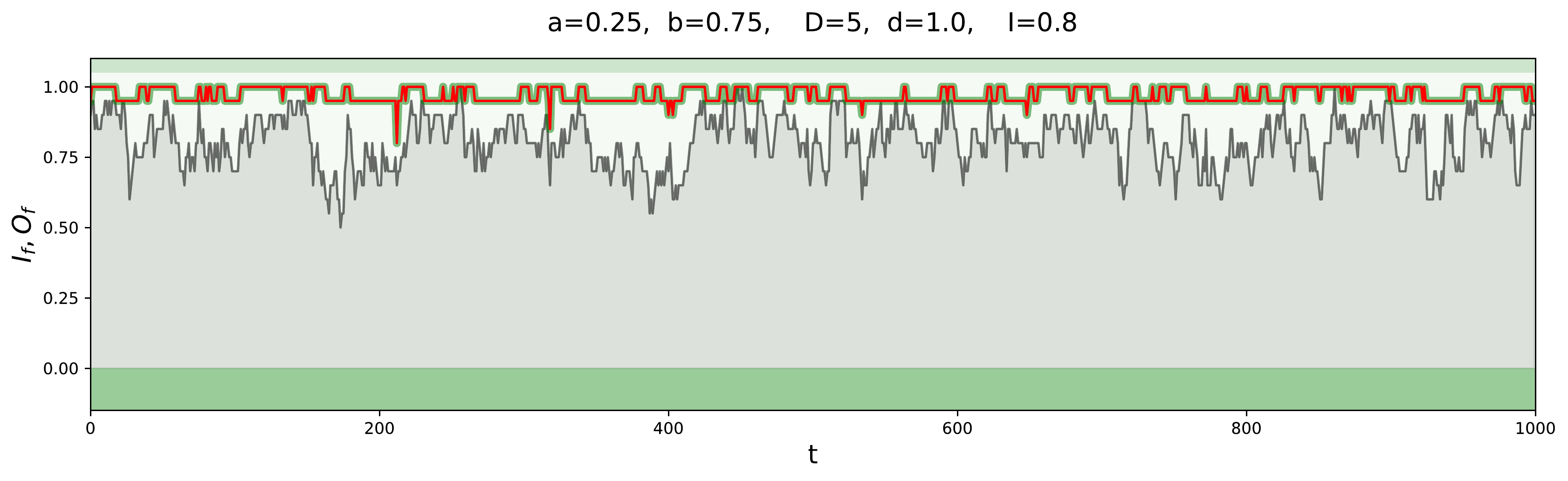}
	\includegraphics[width=1.0\linewidth]{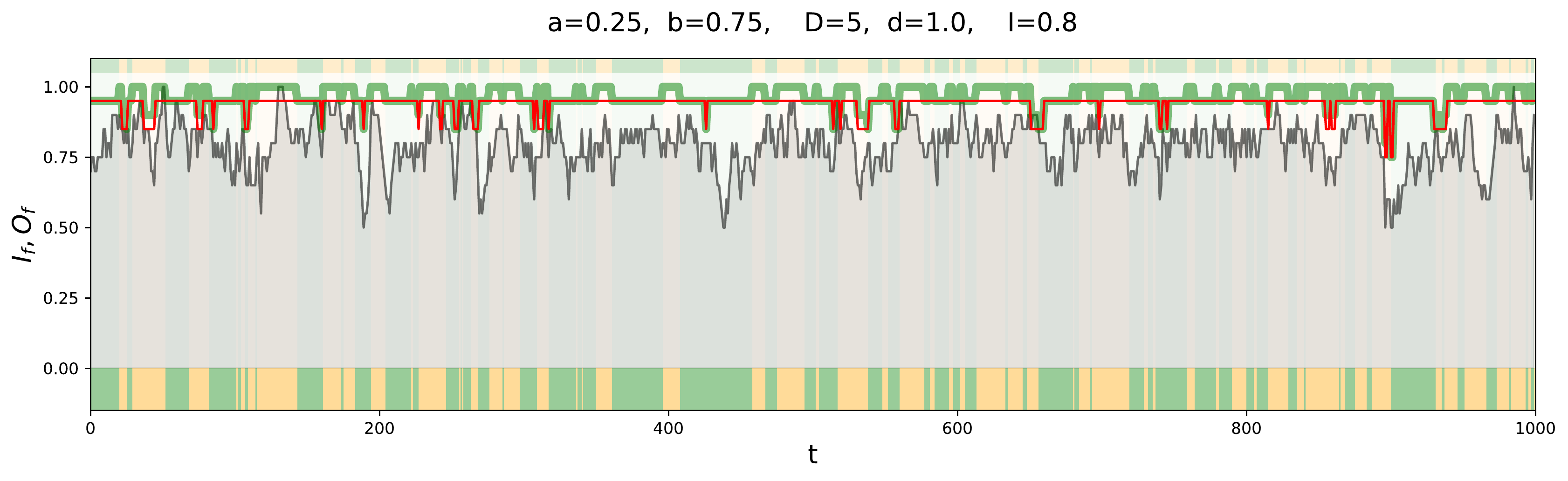}
	\includegraphics[width=1.0\linewidth]{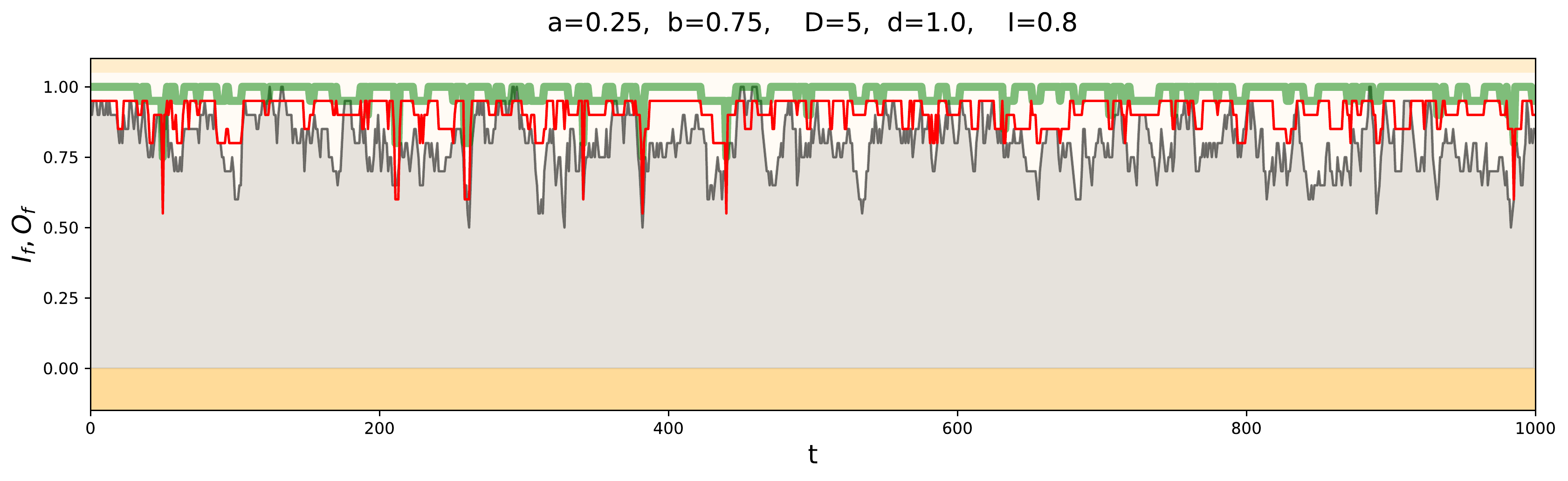}
	\caption[]{\label{fig:trajectories_supplement_1} Three examples of possible time trajectories. From top to bottom: A, AB and B (colour code as defined in Fig.\ref{fig:block_scheme}). Note, that the model parameters are equal. Hence, the differences in the observed disease depend only on the particular realization. Other parameters: $L = 10, H=20$.
	}
\end{figure*}

\begin{figure*}[ht]
	\centering
	\vspace{-3.2cm}
	\includegraphics[width=1.0\linewidth]{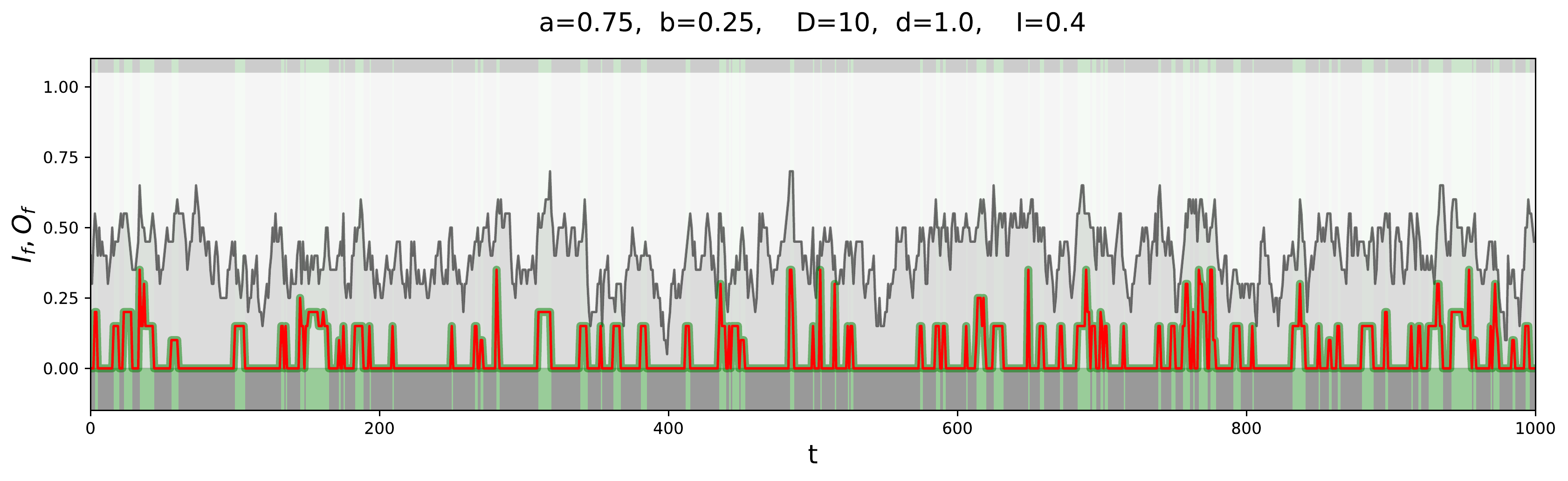}
	\includegraphics[width=1.0\linewidth]{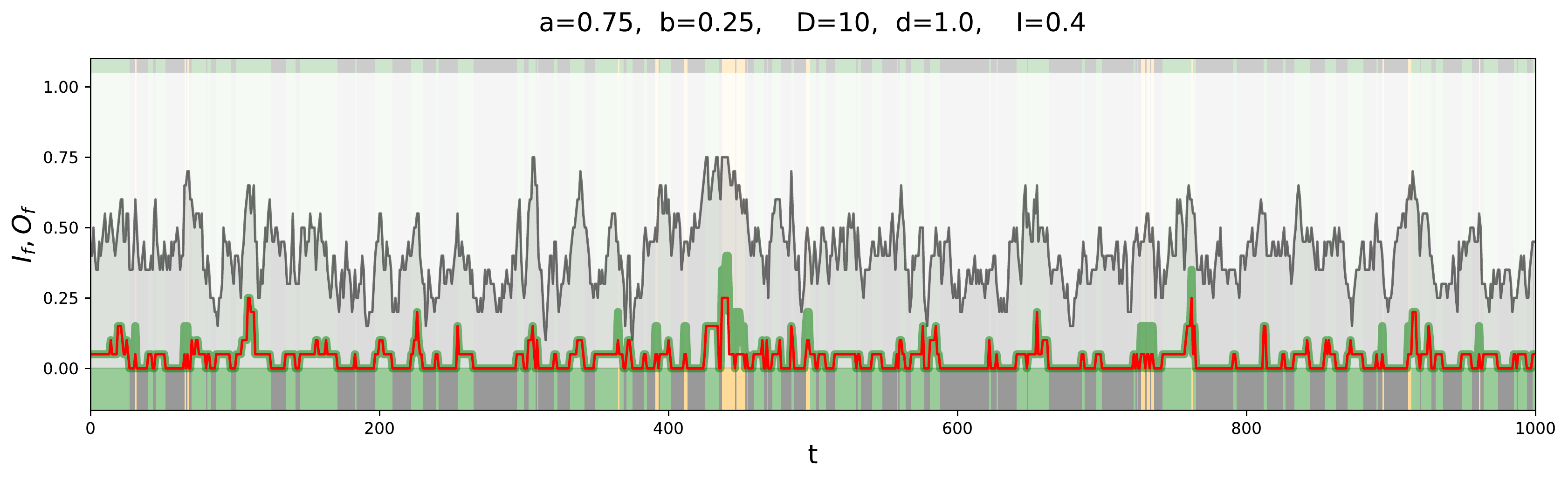}
	\includegraphics[width=1.0\linewidth]{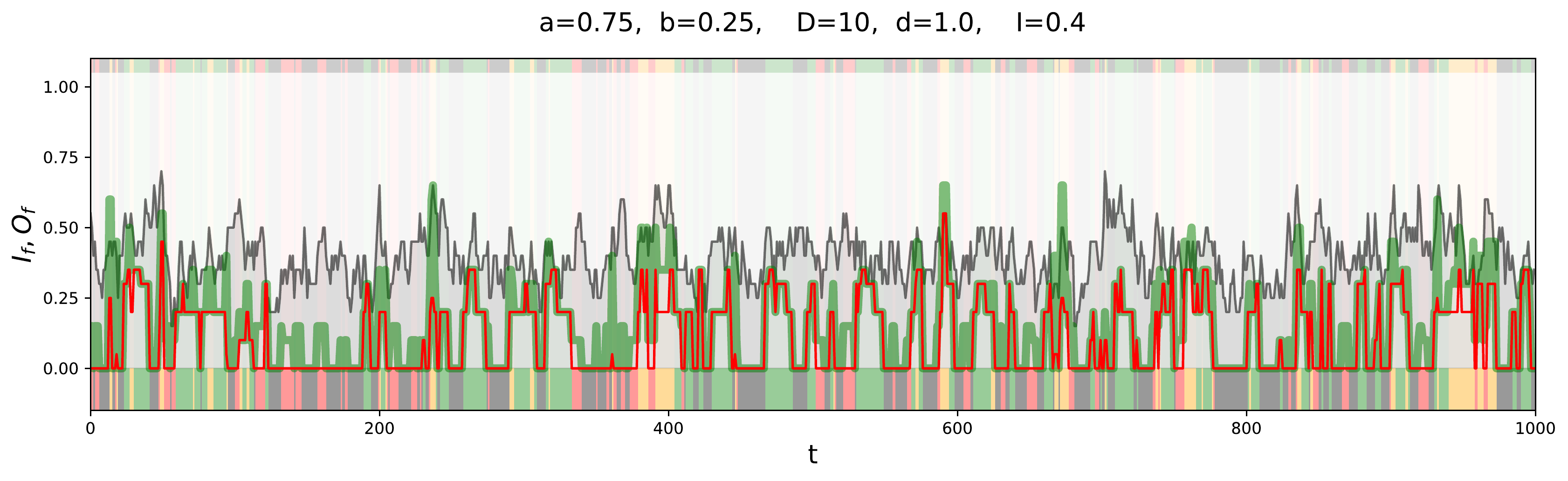}

	\caption[]{\label{fig:trajectories_supplement_2} Three examples of possible time trajectories. From top to bottom: AD, ABD, ABCD (colour code as defined in Fig.\ref{fig:block_scheme}). Other parameters: $L=20$, $H=20$.
	}
\end{figure*}

\begin{figure*}[ht]
	\centering
	\vspace{-2.85cm}
	\includegraphics[width=1.0\linewidth]{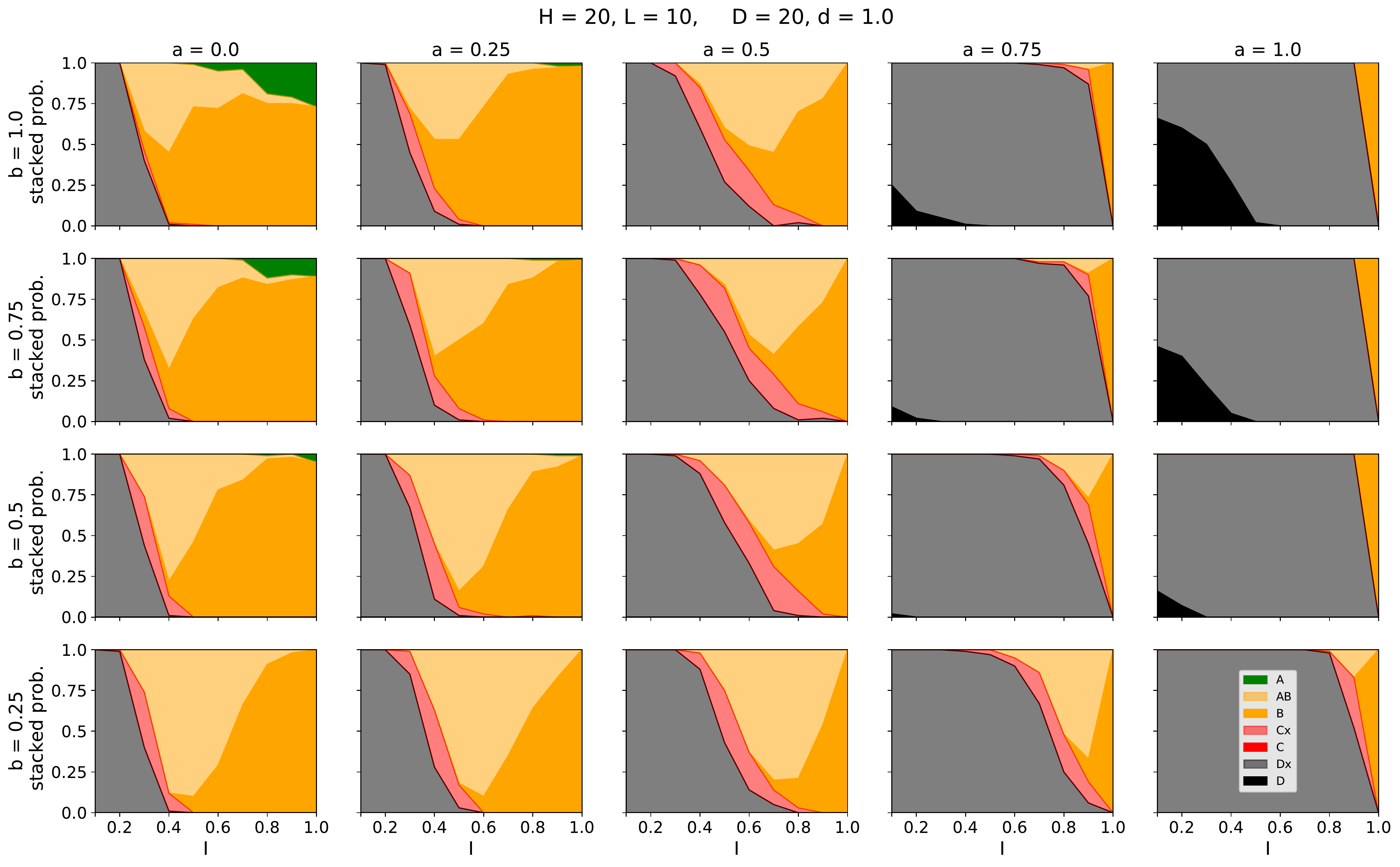}
	\includegraphics[width=1.0\linewidth]{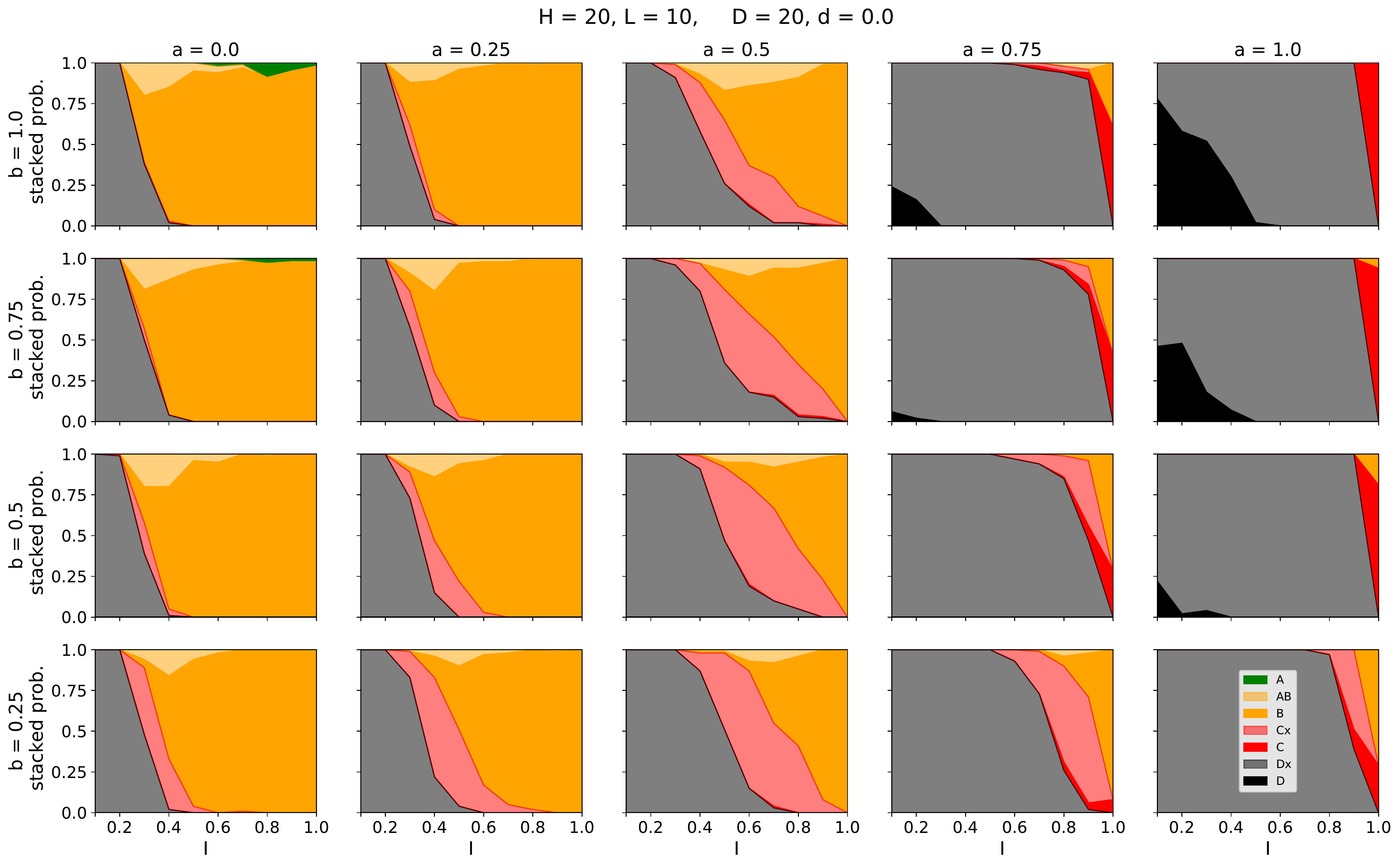}
	\caption[]{\label{fig:dependence_input_supplement_1} Dependence of the observed cases on the fraction of active inputs $\pUp$. For each parameter combination, the system was simulated 100 times for 1000 time steps. The colour code (as defined in Fig.~\ref{fig:block_scheme} a) indicates which cases occurred in the time line.
	}
\end{figure*}

\begin{figure*}[t]
	\centering
	\vspace{-2.85cm}
	\includegraphics[width=1.0\linewidth]{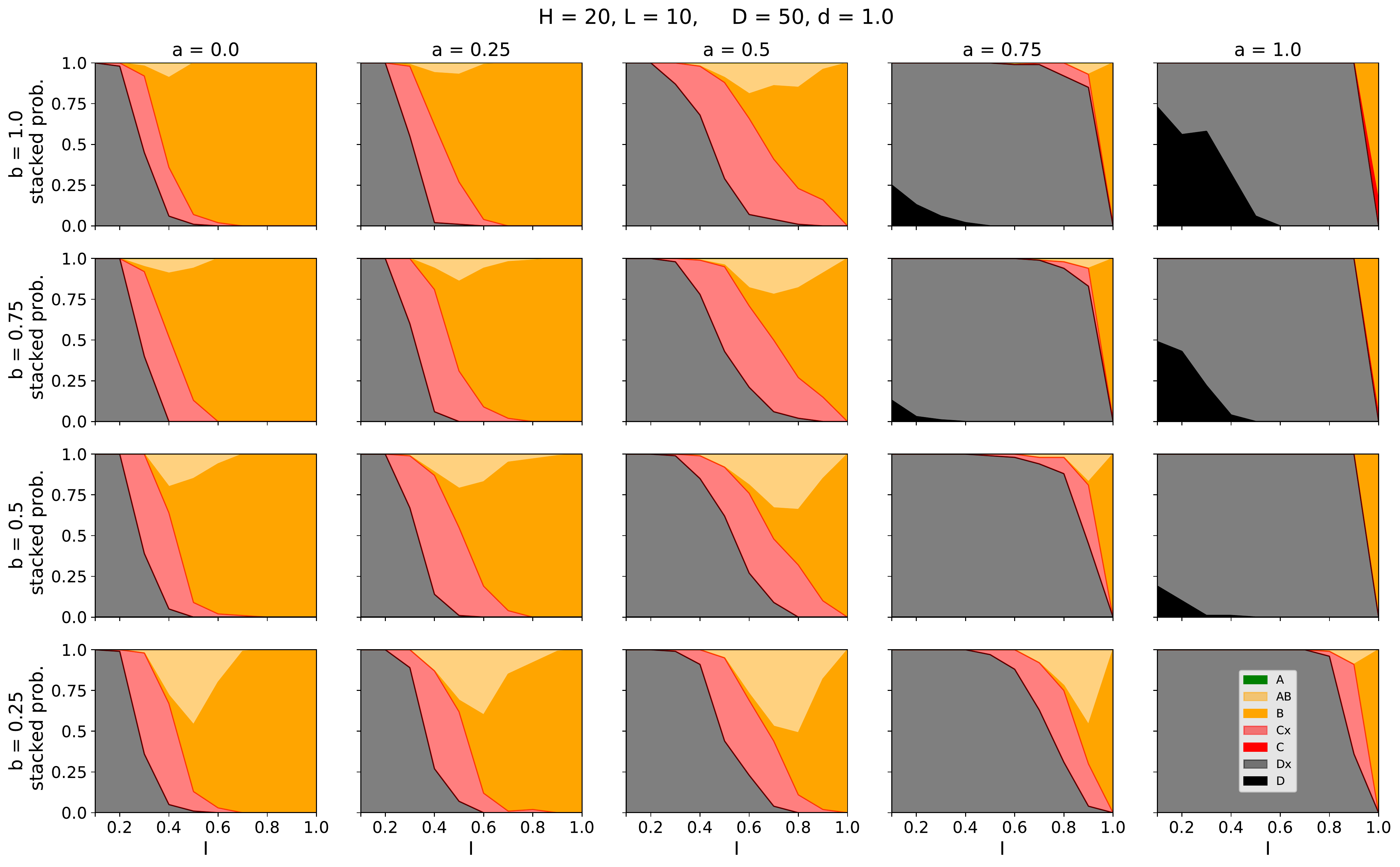}
	\includegraphics[width=1.0\linewidth]{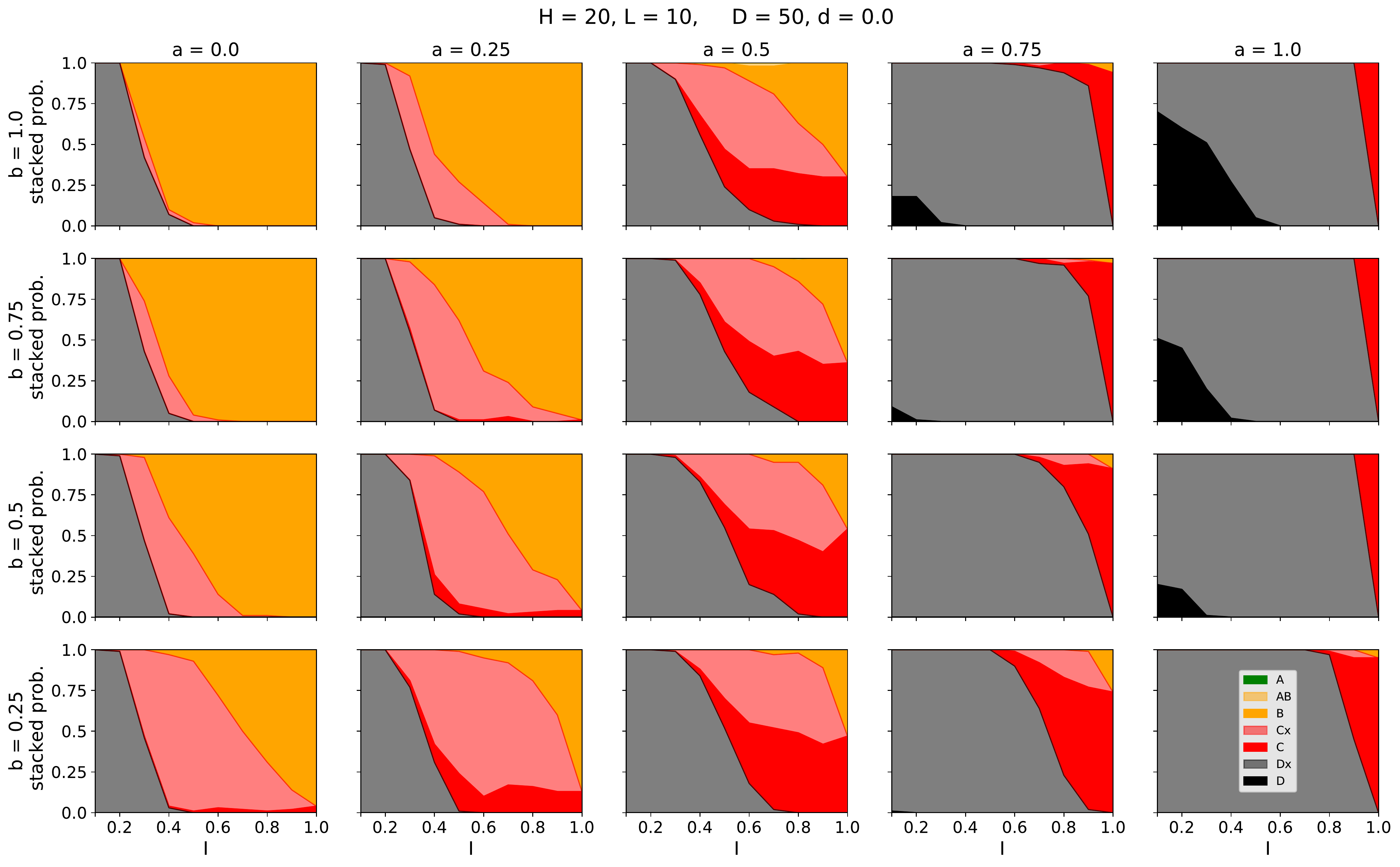}
	\caption[]{\label{fig:dependence_input_supplement_2} Dependence of the observed cases on the fraction of active inputs $\pUp$. For each parameter combination, the system was simulated 100 times for 1000 time steps. The colour code (as defined in Fig.~\ref{fig:block_scheme} a) indicates which cases occurred in the time line.
	}
\end{figure*}

\begin{figure*}[t]
	\centering
	\vspace{-2.85cm}
	\includegraphics[width=1.0\linewidth]{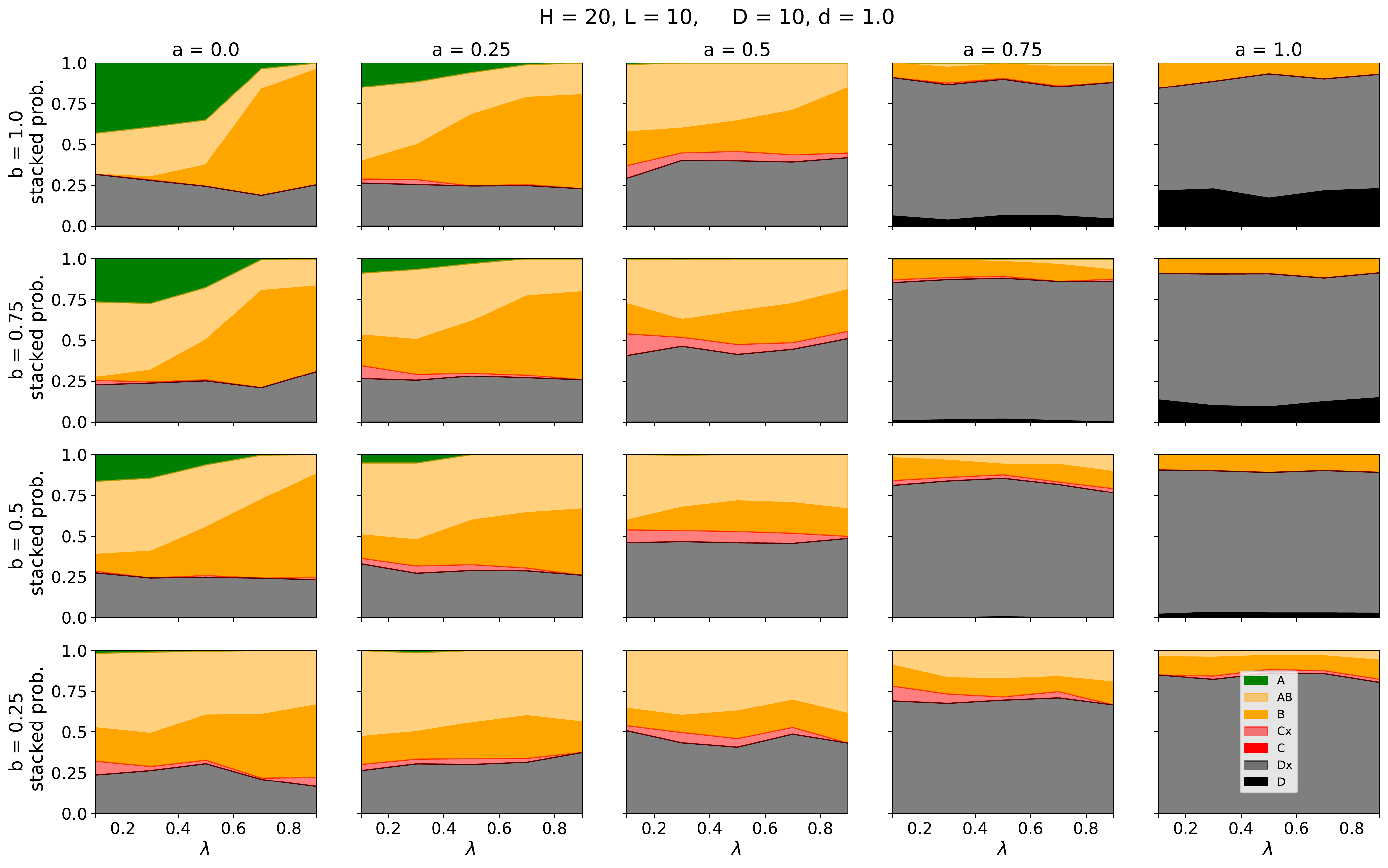}
	\includegraphics[width=1.0\linewidth]{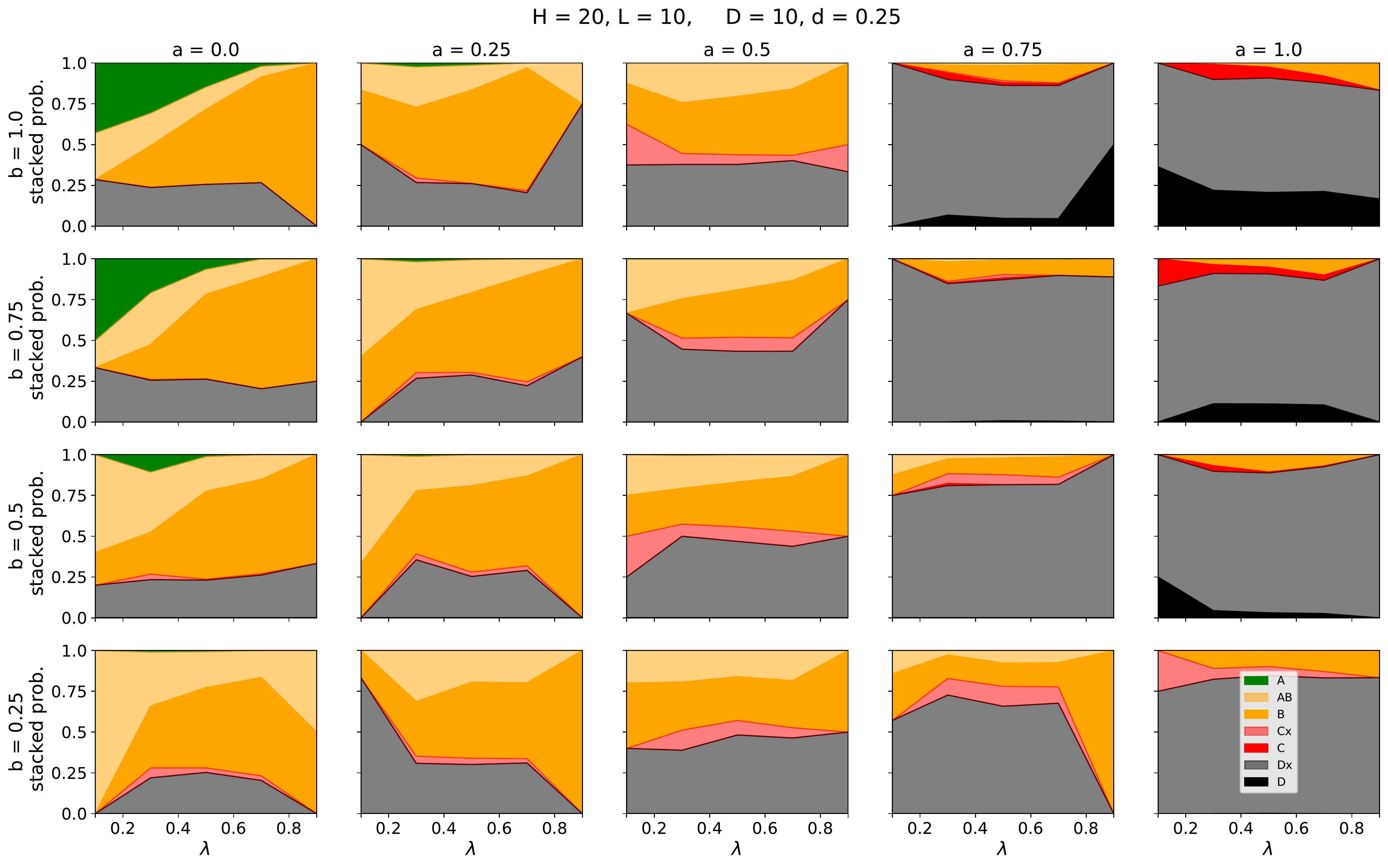}
	\caption[]{\label{fig:dependence_lambda_supplement_1} Dependence of the observed cases on the average position $\disLoc$ of inactive nodes. The colour code (as defined in Fig.~\ref{fig:block_scheme} a) indicates which cases occurred in the time line.
	}
\end{figure*}


\begin{figure*}[t]
	\centering
	\vspace{-2.85cm}
	\includegraphics[width=1.0\linewidth]{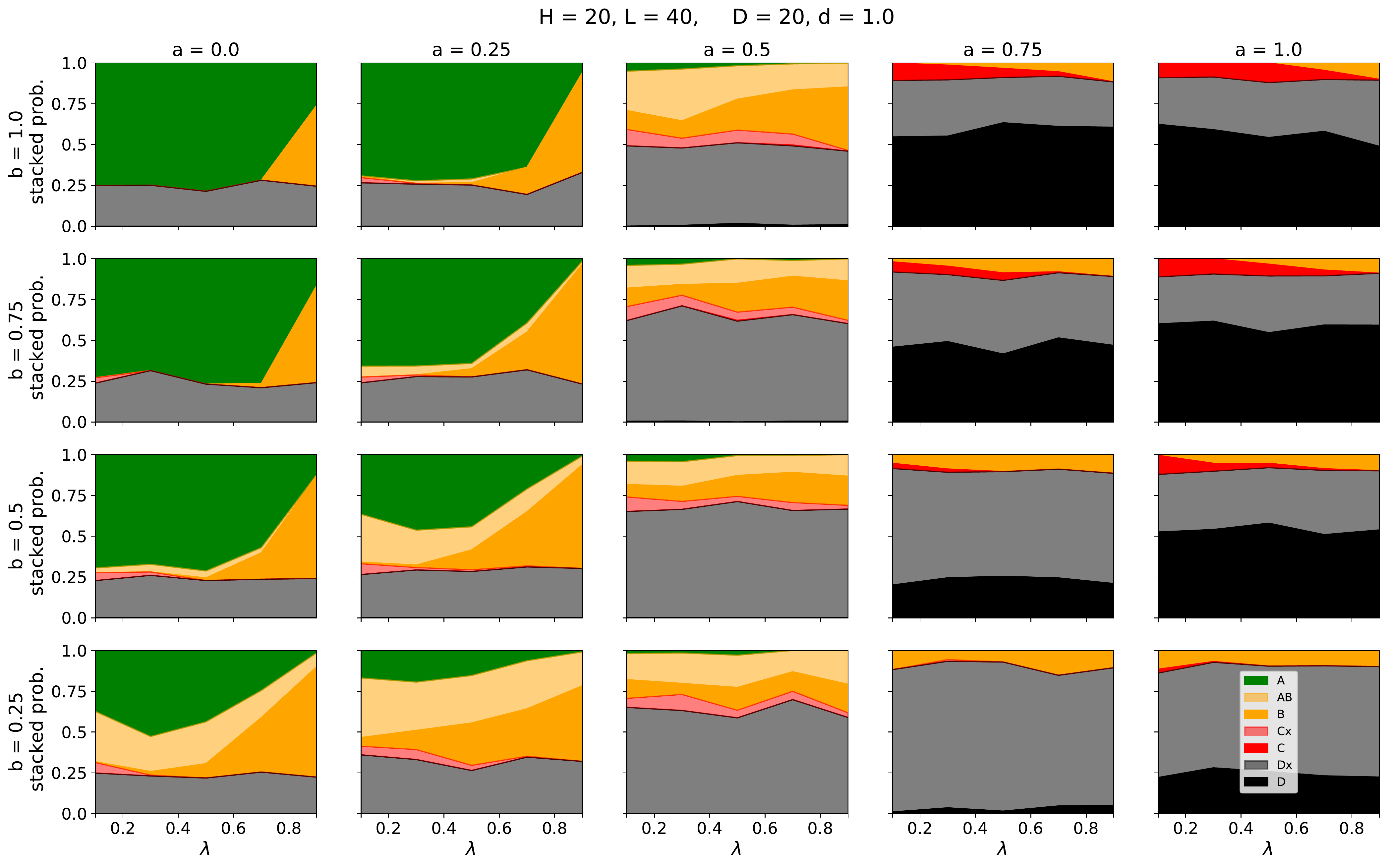}
	\includegraphics[width=1.0\linewidth]{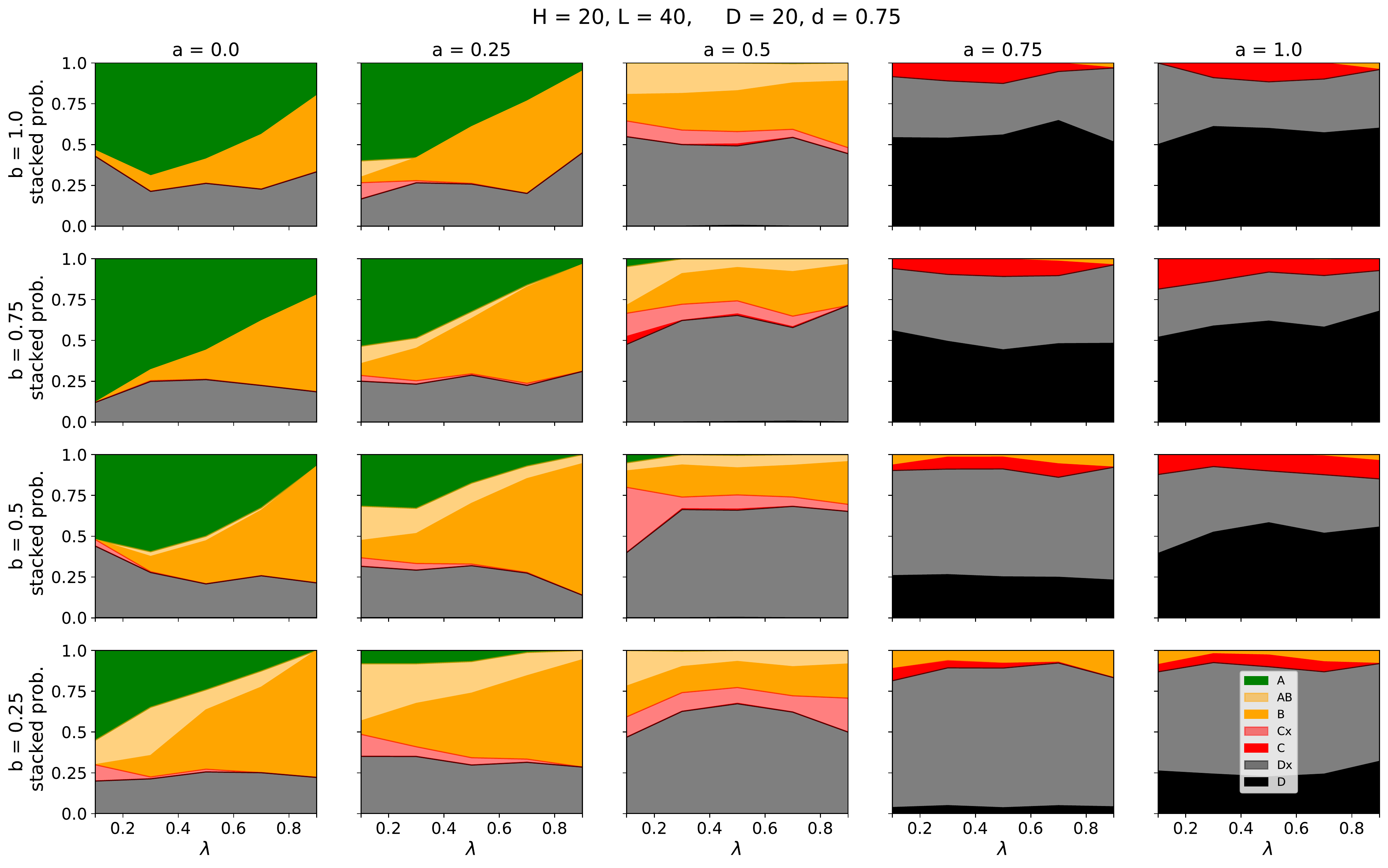}
	\caption[]{\label{fig:dependence_lambda_supplement_2} 
	Dependence of the observed cases on the average position $\disLoc$ of inactive nodes. The colour code (as defined in Fig.~\ref{fig:block_scheme} a) indicates which cases occurred in the time line.
	}
\end{figure*}


\end{document}